\documentclass[10pt,journal,letterpaper,twocolumn,twoside]{IEEEtran} 

\usepackage{amsmath,amsfonts,bm,amssymb,psfrag,ifthen,color,subfigure}
\usepackage{mathrsfs}
\usepackage[justification=centering,font=footnotesize]{caption}
\usepackage{todonotes}
\allowdisplaybreaks[4]

\usepackage[dvips]{epsfig}
\usepackage[dvips]{graphicx}
\usepackage{amsmath,amsfonts,bm,amssymb,psfrag,ifthen,color,subfigure}
\usepackage{algpseudocode}
\usepackage{algorithm}
\usepackage{algorithmicx}
\usepackage{ifthen}
\usepackage{setspace}
\usepackage{cite}
\usepackage{url}
\usepackage{longtable}
\usepackage{stfloats}  
\usepackage{graphics,booktabs,color,subfigure}
\usepackage{float}
\usepackage{diagbox}

\usepackage[dvips]{epsfig}
\usepackage[dvips]{graphicx}
\usepackage{amsmath,amsfonts,bm,amssymb,psfrag,ifthen,color,subfigure,amsthm}
\usepackage{algpseudocode}
\usepackage{algorithm}
\usepackage{algorithmicx}
\usepackage{setspace}
\usepackage{cite}
\usepackage{url}
\usepackage{longtable}
\usepackage{stfloats}  
\usepackage{graphics,booktabs,color,subfigure}
\usepackage{float}
\usepackage{diagbox}
\usepackage{multirow}
\usepackage{epstopdf}

\allowdisplaybreaks[4]

\newcommand{\tabincell}[2]{\begin{tabular}{@{}#1@{}}#2\end{tabular}}

	\usepackage{graphicx}

	\usepackage{graphicx}
	\usepackage{todonotes} 
	
	\usepackage{array}
	\usepackage{hyperref}
	
	\floatname{algorithm}{Algorithm}

\begin{document}
		\title{Feasibility Conditions for Mobile LiFi }
		
		\author{ Shuai~Ma,  Haihong Sheng, Junchang Sun, Hang Li, Xiaodong Liu, Chen Qiu, Majid Safari, Naofal Al-Dhahir and  Shiyin Li
 }
	\maketitle
	\begin{abstract}
		Light fidelity (LiFi) is a  potential key technology  for future 6G networks. However, its feasibility of supporting mobile communications has not been fundamentally discussed. In this paper, we investigate the time-varying channel characteristics of mobile LiFi based on measured mobile phone rotation and movement data. Specifically, we define LiFi channel coherence time to evaluate the correlation of the channel timing sequence.
		Then, we derive the expression of LiFi transmission rate based on the m-pulse-amplitude-modulation (M-PAM). The derived rate expression indicates that mobile LiFi communications is feasible by using at least two photodiodes (PDs) with different orientations.
	    Further, we  propose two channel estimation schemes, and
		 propose a LiFi channel tracking scheme to improve the communication performance.
		 Finally, our experimental results show that the channel coherence time is on the order of tens of milliseconds, which indicates a relatively stable channel.
		In addition, based on the measured data, better communication performance can be realized in the multiple-input multiple-output (MIMO) scenario with a rate of $36 {\rm{Mbit/s}}$, compared to other scenarios. The results also show that the proposed channel estimation and tracking schemes are effective in designing mobile LiFi systems.
	\end{abstract}
	\begin{IEEEkeywords}
Mobile LiFi, coherence time, channel estimation, channel tracking
	\end{IEEEkeywords}
	
	\IEEEpeerreviewmaketitle
	
	\section{Introduction}
	\subsection{Background and Motivation}
	With the rapid development of  information technology, the number of mobile devices are growing exponentially \cite{Forecast_cisco_2019,Zeyu_2021,zhang_2021_statistical}. To relieve the heavy  data traffic load on radio frequency (RF) networks, researchers from both academia and industry are working on new technologies to provide reliable and high-speed wireless data services \cite{Ye_ICCW_2021}. Specifically, light fidelity (LiFi) technology provides ultra-high transmission rates and utilizes high frequency bandwidth to effectively solve the spectrum crisis, which is considered as a key technology for the sixth generation (6G) communications \cite{Haas_2021_ICSJ, Abumarshoud_2022_IVTM}.
	
 To achieve simultaneous lighting and communication, LiFi controls light-emitting diodes (LEDs) to generate light and communication signals, and uses photo-diodes (PDs) to receive the signals \cite{uysal_ICTON_2014,Komine_ICE_2004,Wu_IN_2014,haas_JLT_2015,lee_LCL_2011}. LiFi has several unique features over the traditional wireless fidelity (WiFi) that have been illustrated in existing research \cite{Arfaoui_IJSAC_2021,Zeng_ITC_2020,Ma_TWC_2022}. On the one hand,  LiFi transmission is highly efficient due to the large and unregulated bandwidth of visible light.
 On the other hand, LiFi is virtually immune to interference from other devices and guarantees the data transmission security, thanks to the use of  the visible light band. Moreover, LiFi can be deployed directly through existing LEDs and PDs, making  it easy to build without specialized equipment or infrastructure. Therefore, LiFi technology has a wide application prospect for wireless communication in indoor, airports, hospitals, subways, and other  scenarios \cite{Haas_JOCN_2020}.
	
	However, practical scenarios show that the LiFi channel exhibits dynamics due to variations of user density, blockage, dimming,  and background illumination in the indoor environment \cite{Anwar_ICTON_2019,Lin_WCL_2020,Jiang_IICL_2016}. To this end, there are three important issues to be studied:
	\begin{itemize}
		\item[1)] Determining if LiFi supports mobile communications.
		\item[2)] Exploring influential factors in mobile LiFi communications.
		\item[3)] Investigating channel estimation and  tracking schemes for enhancing the mobile LiFi communication quality.
	\end{itemize}

	\subsection{Related Works}
	Researchers in \cite{Miramirkhani_PTRSA_2020} presented a comprehensive survey for indoor LiFi channel models, by considering the impact of receiver location on indoor mobile users and mobile phone terminals.
	In \cite{Wu_ITC_2020}, the authors found space-time-dependent statistical patterns on channel, bandwidth and outage probability, which  evolve over the environment-confined mobility. In \cite{Miramirkhani_ICL_2017}, the authors moved through different trajectories in a three-dimensional environment with a mobile phone in hand, and constructed a channel model that takes into account the dynamics of indoor LiFi channels. These works indicate that the LiFi channel is indeed affected by the user movements.
	
	In addition, a practical channel model with the terminal rotation (i.e., mobile  terminals scenario) is critical in designing a LiFi system	\cite{   Soltani_IJSAC_2019, Ero_ITC_2019, Soltani_ITC_2019}. In the past, researchers only considered the fixed orientation of mobile devices to simplify the analysis. However, according to the  recent studies, the communication performance of a LiFi system is significantly affected by the device orientation. The authors in \cite{Soltani_IJSAC_2019} studied the statistical characteristics of the signal-to-noise ratio (SNR) caused by the random movement in indoor visible light communication (VLC) networks. Similarly, the authors in \cite{Ero_ITC_2019} investigated the effect of the random orientation and the user location on the channel of the line-of-sight (LOS) condition.  Based on channel models obtained from real measurements, the authors in \cite{Soltani_ITC_2019} conducted a study on the effects of random orientation, user mobility, and obstruction on the SNR and bit error rate for indoor mobile devices. These works fundamentally validate the sensitivity of the LiFi channel to the terminal orientation.
	
	Therefore, it is important to study the time-domain variability of mobile receivers within the LiFi channel \cite{Purwita_TWC_2019,Chen_JLT_2020,bykhovsky_Sensors_2020}. Generally, the coherence time $\footnote{The coherence time indicates that the channels remain essentially unchanged during the coherence time range. However, for conventional RF channels, the coherence time is determined by the Doppler effect. Hence, the equations about coherence time for RF channels can not be applied for mobile LiFi channels.}$  is used to quantify the channel variation characteristics. In \cite{Purwita_TWC_2019}, the authors concluded that the coherence time of the random orientation is in the order of hundreds of milliseconds.  It can be seen that indoor optical wireless communication (OWC) channels are considered to be slowly varying. In addition, the authors in \cite{Chen_JLT_2020} proposed the concepts of the coherence distance and associated angle to directly measure VLC channel variations, which is caused by the receiver movement and rotation. However, this approach ignored the effect of the time factor. In \cite{bykhovsky_Sensors_2020},  based on the coherence time measurement, the authors verified that it is reasonable to assume that the OWC  channel of  a mobile scenario changed slowly for a time period of about 100 milliseconds. However, the experiment neglects the effect of the terminal rotation. 	
	
	For a LiFi system, which supports mobile communication, it is  critical  to estimate accurate channel state information (CSI) \cite{Yaseen_JLT_2021, Hussein_CSPA_2016, Xi_OC_2020, Peng_ICC_2020, Amran_IA_2022}.
	Specifically,  the authors in \cite{Yaseen_JLT_2021} compared various channel estimation approaches for static single-input single-output (SISO)-VLC systems, and showed that the least squares (LS) method is easy to be implemented. However, the performance of these approaches is inadequate due to the lack of prior channel statistics information.
	The authors in \cite{Hussein_CSPA_2016} evaluated
	LS and minimum mean square error (MMSE) channel estimation algorithms for indoor orthogonal frequency division multiplexing (OFDM) VLC systems, and showed that the MMSE has better performance at higher SNRs.
	The authors in \cite{Xi_OC_2020} investigated channel estimation  by using deep neural networks,  which only considered LOS channels.
	For mobile scenarios, channel tracking methods can be adopted to evaluate the CSI. In \cite{Peng_ICC_2020}, the authors proposed a channel tracking scheme based on long short term memory (LSTM) model to compensate for the negative effects of the imperfect CSI. The results showed that the system secrecy performance in high mobility scenarios can be improved based on the LSTM-based channel tracking method.
	However, studies on exploring the impact of the mobile LiFi systems on mobile communication with measurement data are limited, and  investigation of the practical time-varying features of the mobile LiFi channel is ignored.

	\subsection{Contributions}
	Given the discussions above, we aim to obtain the channel characteristics based on practical measurement experiments, and then show that  mobile LiFi  is feasible in multiple-input multiple-output (MIMO) scenarios. Moreover, we establish systematic channel estimation and tracking procedures. 	
	The main contributions of this work are summarized as follows:	
	\begin{itemize}
		\item
		First,  we study the timing channel characteristics and derive the coherence time expression of  mobile LiFi systems  based on experimental data. Our results show that the mobile LiFi channel has a time-varying feature, and the channel variations and the coherence time are related to the person's posture, mobile phone orientation,  PD position, user location, and walking speed. Moreover, the results show that  the channel coherence time is on the order of tens of milliseconds, which indicates that the LiFi channel is stable during this period.
		
		\item
		 Then, we study the effects of the number of LEDs and PDs in the walking and sitting scenarios. Based on the considered MIMO mobile LiFi scenarios, we derive an expression for the achievable data rate in the mobile LiFi scenario to evaluate the system performance.
		Experimental results show that multiple PDs  can support a data rate of $36 {\rm{Mbit/s}}$,  and the application of multiple LEDs improves the performance even further.
		
		\item		
		To enhance mobile LiFi communication performance, two channel estimation schemes are proposed: channel  estimation coding and deep learning channel estimation. By comparing their performance with the LS scheme, we show that the channel  estimation coding and deep learning channel estimation scheme have better performance. The deep learning channel estimation scheme achieves the best performance in mobile LiFi channel estimation. Meanwhile, to obtain more accurate CSI, the long short term memory (LSTM) scheme is used to track the channel time-varying characteristics. Our results show that the LSTM model has better tracking performance by comparing with the conventional recurrent neural network (RNN) scheme.
	\end{itemize}

	The rest of this paper is organized as follows. The system
	model is presented in Section II. Section III presents the
	mobile LiFi timing features. Section IV studies the communication performance of mobile LiFi. Sections VI and V investigate channel estimation and tracking of mobile LiFi, respectively.   Finally, Section VII concludes this paper. In addition, Table \ref{tableI} summarizes the main acronyms  and	 Table \ref{tableII} summarizes the relevant parameters  in  this paper.
	
	\emph{Notations}:  Vectors and matrices are represented by
	boldfaced lowercase and uppercase letters, respectively.
	The notations ${\left(  \cdot  \right)^{\rm{T}}}$, {$\mathbb{E} \left[  \cdot  \right]$},  ${\rm{Tr}}\left(  \cdot  \right)$ and ${\left(  \cdot  \right)^\dag }$ represent
	the transpose, expectation,  trace and pseudo-inverse of a matrix, respectively.
	$\otimes$ represents the element-wise multiplication, $\oplus $ represents the matrix addition, $\delta{\left(  \cdot  \right)}$ is impulse function, and $\varepsilon{\left(  \cdot  \right)}$ is step function.

	\begin{table}[h]
		\caption{Summary of Main Acronyms}
		\centering
		\label{tableI}
		\begin{tabular}{|m{2cm}|m{6cm}<{\centering}|}
			\hline
			\multicolumn{1}{|c|}{Notation}    & Description \\
			\hline
			LiFi & Light Fidelity  \\
			\hline
			PD &  Photo Diode  \\
			\hline
			LED & Light-Emitting Diodes  \\
			\hline
			UE & User Equipment  \\
			\hline
			LOS & Line Of Sight  \\
			\hline
			FOV & Filed Of View  \\
			\hline
			CDF & Cumulative Distribution Function  \\
			\hline
			LS & Least Squares  \\
			\hline
			CDRN & Convolutional neural network based Deep Residual Network  \\
			\hline
			NMSE & Normalized Mean-Squared Error  \\
			\hline
			LSTM & Long Short-Term Memory Error  \\
			\hline
		\end{tabular}
	\end{table}

	\begin{table}[H]
		\caption{Summary of Key Notations}
		\label{tableII}
		\centering
		\begin{tabular}{|m{2cm}|m{6cm}<{\centering}|}
			\hline
			\rule{0pt}{8pt}Notation  & Description   \\ \hline
			\rule{0pt}{7.5pt}${{\bf{u}}_l}$ &  \tabincell{c}{LED position} \\ \hline
			\rule{0pt}{7.5pt}${{\bf{u}}_u}$ &  \tabincell{c}{UE centroid position}\\ \hline
			\rule{0pt}{7.5pt}${{\bf{u}}_{p}}$ &  \tabincell{c}{ PD position}\\ \hline
			\rule{0pt}{7.5pt}${\bf{R}}{\left( t \right)}$ &  Rotation matrix \\ \hline
			\rule{0pt}{7.5pt}${{\bf{n}}_l}$ &  LED rotation \\	\hline		
			\rule{0pt}{7.5pt}${{\bf{n}}_p}$ &  PD rotation \\	\hline	
			\rule{0pt}{7.5pt}$\rm{v}$ &  Walking speed \\ 	\hline		
			\rule{0pt}{7.5pt}${\rho _{L}}$ &  Normalized autocorrelation coefficient  \\ 	\hline	
			\rule{0pt}{7.5pt}${T_c}$ &  coherence time  \\ 	\hline	
			\rule{0pt}{7.5pt}$R_1$ &  Achievable data rate of the SISO scenario \\ 	\hline	
			\rule{0pt}{7.5pt}$R_2$ &  Achievable data rate of the SIMO scenario \\ 	\hline	
			\rule{0pt}{7.5pt}$R_3$ &  Achievable data rate of the MISO scenario \\ 	\hline	
			\rule{0pt}{7.5pt}$R_4$ &  Achievable data rate of the MIMO scenario \\ 	\hline	
			\rule{0pt}{7.5pt}$\Delta h$ & Normalized channel error  \\ 	\hline	
			
		\end{tabular}
	\end{table}

	\section{System Model }
	\begin{figure}[htbp]
		\centering
		\includegraphics[width=8.5cm]{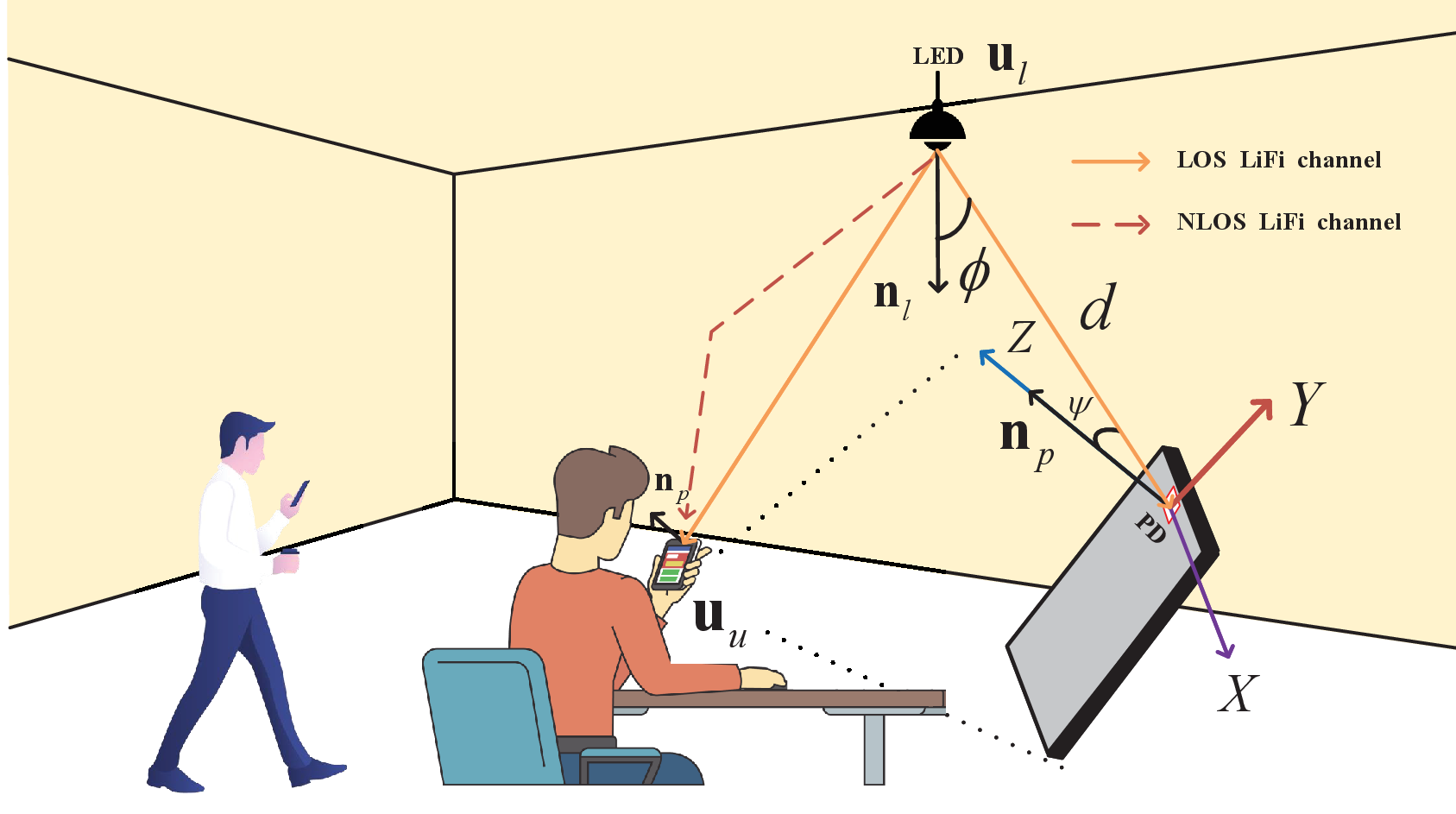}
		\caption{Mobile LiFi System}
		\label{img1}
	\end{figure}	
	Consider a mobile LiFi system as shown in Fig. $\ref{img1}$. The system consists of a ceiling mounted LED and a user equipment (UE). The LED is vertically downwards for lighting. The UE is equipped with PDs and an inertial measurement unit (IMU) is randomly located in the room (sitting user with   red clothes or walking user with white clothes).
	
	To better describe the  location information, we establish a three-dimensional coordinate system with the origin located at the corner of the room. Denote the LED position as ${{\bf{u}}_l} = {\left[ {{x_l},{y_l},{z_l}} \right]^{\rm{T}}}$, the UE centroid position as ${{\bf{u}}_u} = {\left[ {{x_u},{y_u},{z_u}} \right]^{\rm{T}}}$, and the  PD position as ${{\bf{u}}_{p}} = {\left[ {{x_{p}},{y_{p}},{z_{p}}} \right]^{\rm{T}}}$. Since the LED direction is vertically downward, the LED normal vector can be expressed as ${{\bf{n}}_l} = {\left[ {0,0, - 1} \right]^{\rm{T}}}$.
	We use $\alpha $, $\beta $, and $\gamma $ to represent rotation angles relative to  the $Z$, $X$, and $Y$ axes, respectively. The rotation matrix ${\bf{R}}{\left( t \right)}$ at  time $t$ can be expressed as
	\begin{align}
		{\bf{R}}{\left( t \right)} = {{\bf{R}}_\alpha }{\left( t \right)}{{\bf{R}}_\beta }{\left( t \right)}{{\bf{R}}_\gamma }{\left( t \right)},
	\end{align}
	where ${{\bf{R}}_\alpha }{\left( t \right)}$, ${{\bf{R}}_\beta }{\left( t \right)}$, and ${{\bf{R}}_\gamma }{\left( t \right)}$ are, respectively, given by
	\begin{subequations}
		\begin{align}
			{{\bf{R}}_\alpha }{\left( t \right)} = \left[ {\begin{array}{*{20}{c}}
					{\cos \alpha }&{ - \sin \alpha }&0\\
					{\sin \alpha }&{\cos \alpha }&0\\
					0&0&1
			\end{array}} \right],\\
			{{\bf{R}}_\beta }{\left( t \right)} = \left[ {\begin{array}{*{20}{c}}
					1&0&0\\
					0&{\cos \beta }&{ - \sin \beta }\\
					0&{\sin \beta }&{\cos \beta }
			\end{array}} \right],\\
			{{\bf{R}}_\gamma }{\left( t \right)} = \left[ {\begin{array}{*{20}{c}}
					{\cos \gamma }&0&{\sin \gamma }\\
					0&1&0\\
					{ - \sin \gamma }&0&{\cos \gamma }
			\end{array}} \right].
		\end{align}
	\end{subequations}

	The PD normal vector ${{\bf{n}}_p} {\left( t \right)}$ after the rotation can be expressed as
	\begin{align}\label{n}
		{{\bf{n}}_p} {\left( t \right)}= {\bf{R}}{\left( t \right)}{{\bf{n}}_{p,0}},
	\end{align}
	where ${\bf{n}}_{p,0}$ is the initial normal PD vector.
	The PD position ${{\bf{u}}_p}{\left( t \right)}$ after the rotation  can be expressed as
	\begin{align}\label{u}
		{{\bf{u}}_p}{\left( t \right)} = {{\bf{u}}_u} + {{\bf{r}}_p}{\left( t \right)},
	\end{align}
	where ${{\bf{r}}_p}{\left( t \right)} = {\bf{R}}{\left( t \right)}\left( {{{\bf{u}}_{p,0}} - {{\bf{u}}_u}} \right)$ represents the position deviation of the  PD relative to the UE  center point, and ${\bf{u}}_{p,0}$ represents the initial position of PD.
		
	\subsection{LiFi Channel}	
	In the  considered system,  the channel gain $h\left(t\right)$ can be written as a superposition \cite{Jungnickel_JSAC_2002}, that is
	\begin{align}\label{ht}
		{h}\left(t\right) = {h_{{\rm{LOS}}}}\delta \left(t\right) + {h_{{\rm{NLOS}}}}\left(t - \Delta {t_{{\rm{NLOS}}}}\right),
	\end{align}	
	where ${h_{{\rm{LOS}}}}\delta \left(t\right)$ represents the LOS link gain, ${h_{{\rm{NLOS}}}}\left(t - \Delta {t_{{\rm{NLOS}}}}\right)$ represents the diffuse reflections link gain, $ \Delta {t_{{\rm{NLOS}}}}$ describes the delay between the LOS signal and the onset of the diffuse signal.
	
	The LOS link gain ${ {h_{{\rm{LOS}}}}}$ between the LED and  PD can be modeled as a Lambertian model of order $m$ \cite{Liu_ITVT_2019}, given as
	\begin{align}
		{h_{{\rm{LOS}}}} = \left\{ {\begin{array}{*{20}{c}}
				{\frac{{\left(m + 1\right){A_{\rm PD}}}}{{2\pi d^2}}{g_f}{{\cos }^m}{\phi }\cos {\psi }},&{0 \le {\psi } \le {\Psi _{{\rm{FOV}}}},}\\
				0,&{{\psi } \ge {\Psi _{{\rm{FOV}}}},}
		\end{array}} \right.
	\end{align}
	where ${A_{\rm PD}}$ is the receiving area of the  PD; ${g_f}$ is the gain of the optical concentrator;  ${\Psi _{{\rm{FOV}}}}$ represents the field of view (FOV) of the receiver;  ${d}$ is the Euclidean distance between LED and the  PD; ${\phi }$ and ${\psi }$ are the exit angle and incidence angle from the LED to  PD, respectively.	
	Based on \eqref{n} and \eqref{u}, ${ {h_{{\rm{LOS}}}}}$ can be further expressed as
	\begin{figure*}[h]
		\begin{equation}\label{h_Los}
			{h_{{\rm{LOS}}}} = \left\{ {\begin{array}{*{20}{c}}
					{\frac{{\left( {m + 1} \right){A_{{\rm{PD}}}}{g_{\rm{f}}}{{\left( {{z_l} - {z_p}} \right)}^m}{{\left( {{{\bf{u}}_l} - \left( {{{\bf{u}}_u} + {{\bf{r}}_p}} \right)} \right)}^{\rm{T}}}{{\bf{n}}_p}}}{{2\pi {{\left\| {{{\bf{u}}_l} - \left( {{{\bf{u}}_u} + {{\bf{r}}_p}} \right)} \right\|}^{m + 3}}\left\| {{{\bf{n}}_p}} \right\|}},}&{0 \le {\psi } \le {\Psi _{{\rm{FOV}}}},}\\
					0,&{{\psi } \ge {\Psi _{{\rm{FOV}}}}}.
			\end{array}} \right.
		\end{equation}
	\end{figure*}
	
	The channel gain of the first reflected ray between the LED and PD within the FOV is given by \cite{Jungnickel_JSAC_2002}
	\begin{align}\label{h_NLOS}
		{h_{{\rm{NLOS}}}}(t - \Delta {t_{{\rm{NLOS}}}}) &=2\pi {f_c}\frac{{{A_{\rm PD}}{\rho _{{\rm{NLOS}}}}}}{{{A_r}(1 - {\rho _{{\rm{NLOS}}}})}}\exp \left( { - 2\pi f_c} \right.\\ \nonumber
		& \times  \left. {\left( {t - \Delta {t_{{\rm{NLOS}}}}} \right)} \right)\varepsilon \left( {t - \Delta {t_{{\rm{NLOS}}}}} \right),
	\end{align}
	where ${\rho_{\mathrm{NLOS}}}$ is the reflection coefficient; $A_r$ is the area of the indoor scene surface; $ {f_c}$ is the cutoff frequency.

	According to \eqref{ht}, \eqref{h_Los}, \eqref{h_NLOS}, it is noted that the LiFi channel state is time-varying. For the considered moving  LiFi scenarios, the LiFi channel is dynamically changing due to the randomness of the PD orientation and position.

	\subsection{Coherence Time of LiFi}

	The coherence time refers to the characteristic time during which the channel remains relatively stable. This metric is closely related to UEs' movements.
	
	Based on the measured data, the channel gain  $h\left[ n \right] $ is modeled by a stationary discrete-time continuous-valued stochastic  process.
	The autocorrelation ${C_{h\left[ n \right]}}$ of the channel gain $h\left[ n \right]$ can be expressed as
	\begin{align}
		\begin{split}
			{C_{h\left[ n \right]}} ={\mathbb{E}} \left[ {\left( {h\left[ n \right] - {{\overline {{h}} }} } \right)\left( {h\left[ {n + L} \right] - {{\overline {{h}} }} } \right)} \right],
		\end{split}
	\end{align}
	where $L$ represents the time slot and $ {\overline {{h}} } $ represents the LiFi channels mean value. 
	
	In addition, the correlation function of the channel is normalized to provide a fair measure of the LiFi channel transformation in different scenarios. The normalized autocorrelation coefficient ${\rho _{L}}$ is defined as
	\begin{align}
		\begin{split}
			{\rho_{L} } =  \frac{\mathbb{E}{\left[ {\left( {h\left[ n \right] - {{\overline {{h}} }} } \right)\left( {h\left[ {n + L} \right] - {{\overline {{h}} }} } \right)} \right]}}{{{\mathbb{E}} \left[ {\left( {h\left[ n \right] - {{\overline {{h}} }} } \right)^2} \right]}}.
		\end{split}
	\end{align}
	
	The coherence time ${T_c}$ is the interval where ${\rho_L }$ is less than a threshold ${{\eta _{{\rm{th}}}}}$, i.e., the coherence time ${T_c}$ satisfies
	\begin{subequations}
		\begin{align}
			{T_{{\rm{c}}}} &= {n_{{\rm{c}}}}\Delta t,\\
			{n_{{\rm{c}}}} &= \mathop {\arg }\limits_k \left\{ {{\rho_L }\left(k\right) = {\eta _{{\rm{th}}}}} \right\},
		\end{align}
	\end{subequations}
	where  $\Delta t = 1/{f_s}$ and ${f_s}$ is the sampling frequency.

	\section{Experimental Study on Channel coherence time of Mobile LiFi}
	
	\subsection{Experimental Setup}
	
\begin{figure}[tb]
		\centering
		\includegraphics[width=8cm]{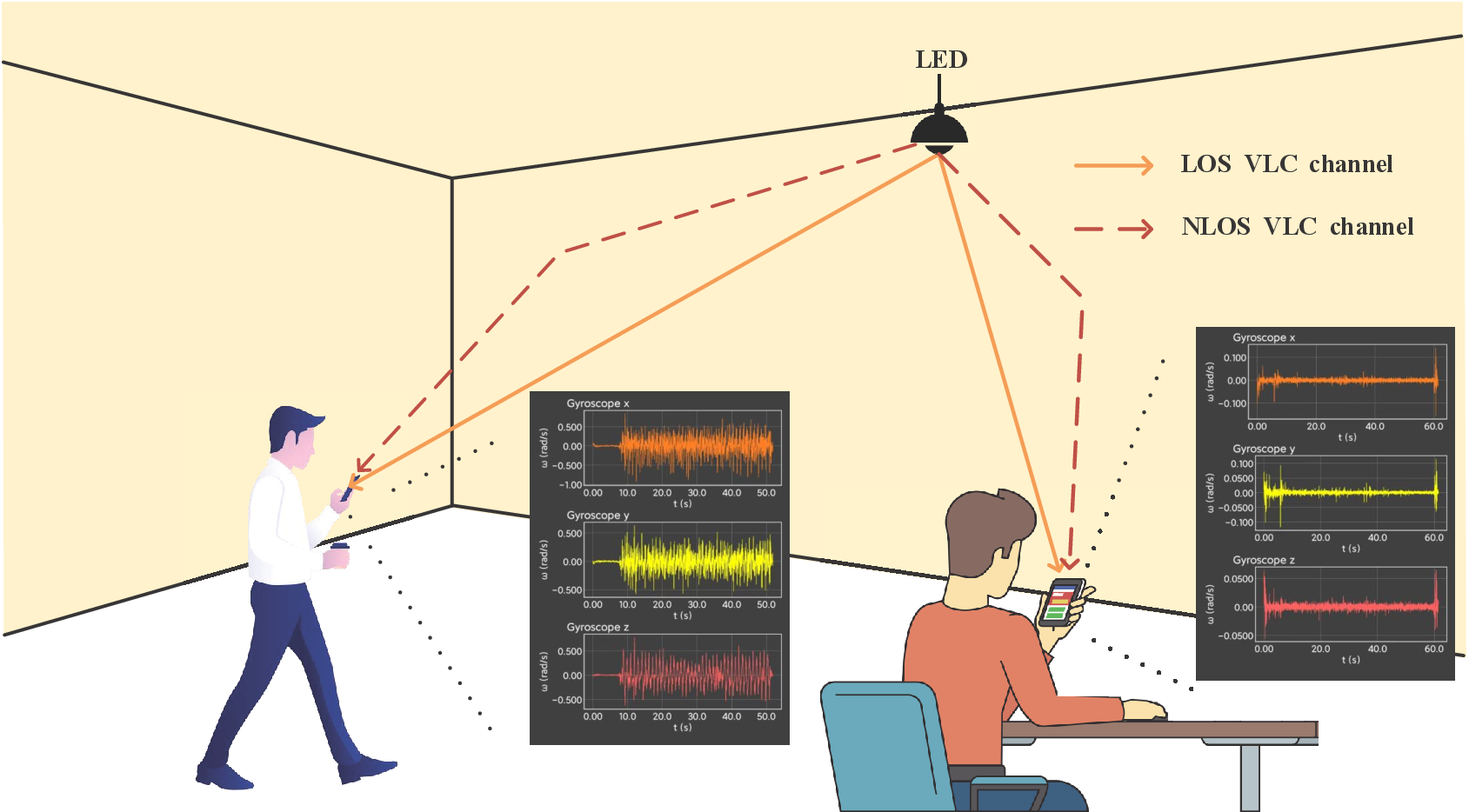}
		\caption{Measurement Scenario.}
		\label{measure}
	\end{figure}

	We conduct experiments indoors,
	where a LED is deployed at the center of the ceiling and $3 {\rm{m}}$ above the floor for the data
	transmission. {The dimensions of the experimental mobile phone are $15.3 {\rm{cm}} \times  7.2 {\rm{cm}} \times 0.85 \rm{cm}$.} In addition, we organized 50 testers  to participate in the experiments, where we separately considered sitting and walking scenarios.
	An illustration of the experimental scenario  is shown in  Fig. \ref{measure}.
	
	Specifically, we extract the measurement data from the phone gyroscope, and record the instantaneous rotation angular velocity of the phone through the software (Phyphox). {After a certain time of acquisition, the data is exported and processed through MATLAB.} The main measured values of the sensor include the time index and the angular rotating velocity in ${\rm{rad/s}}$.
	The specific experimental steps are summarized as follows:
	
	\subsubsection{Sitting Scenario}
	
	\begin{itemize}
		\item {All testers sit in a $4 {\rm{m}} \times  4 {\rm{m}}$ room with a distance space of  $0.5 {\rm{m}}$.}
		\item {Every tester uses the mobile phone for $180 {\rm{s}}$ with the test software running.}
		\item  {Every tester repeats measurements  50 times.
		}
	\end{itemize}

	\subsubsection{Walking Scenario}
	\begin{itemize}
		\item {Every tester walks in a straight line along a $2 {\rm{m}}\times 100 {\rm{m}}$ corridor.
		}
		\item {During the walking process, the tester keeps the test software running and uses the mobile phone normally.}
		\item  {Every tester repeats measurements 50 times.}
	\end{itemize}

	During the experiment, each tester strictly follows the experimental steps to simulate daily mobile phone usage behavior (browsing the web, using the mobile phone in portrait or horizontal mode and etc.). To obtain an accurate rotation angle of the phone, testers are required to place the phone horizontally for $3\rm{s}$ before the experiment to obtain a relatively stable initial value.

	\subsection{Experimental Results}
\begin{figure}[htbp]
		\centering
		\includegraphics[width=8.5cm]{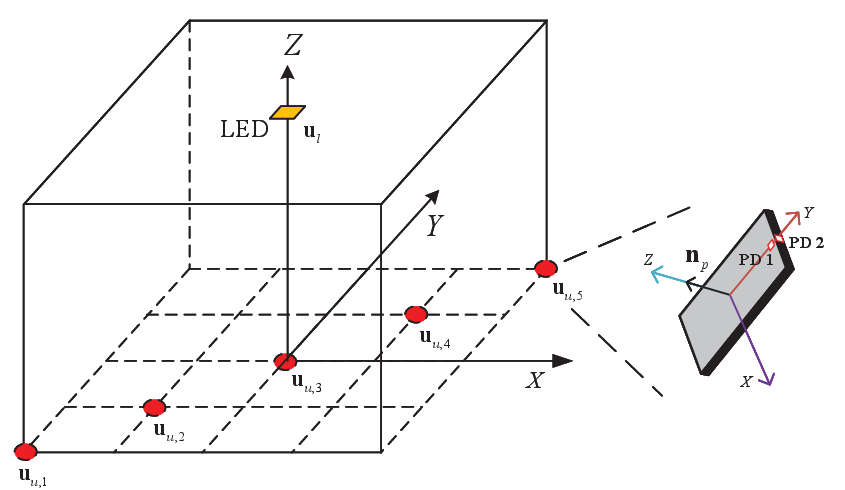}
		\vskip-0.2cm\centering {\footnotesize}
		\caption{Room distribution map}
		\label{map}
	\end{figure}

	In our experiments, we also consider the effect of the PD position on channel characteristics. We used two PDs with different positions to collect the data. Specifically, the initial normal vector of the first PD (PD1) is set as  ${{\bf n}_{p1,0}} = {\left[ {0,0,1} \right]^{\rm{T}}}$ and the initial normal vector of the second PD (PD2) is set as ${{\bf n}_{p2,0}} = {\left[ {0,1,0} \right]^{\rm{T}}}$. The specific PD locations are shown in Fig. \ref{map}, and the detailed
	experiment parameters are given in Table \ref{tableii}.	
	Moreover, the original data is disclosed at the following URL: $\rm{\href{https://pan.baidu.com/s/1S7oFC01q8I84IUhZXuMy2Q}{\textcolor{blue}{https://pan.baidu.com/s/1S7oFC01q8I84IUhZXuMy2Q}}}$, with the extraction code of CUMT.
	\begin{table}[h]
		\caption{Experiment Parameters}
		\label{Simulation1}
		\centering
		\label{tableii}
		\begin{tabular}{|m{6cm}|m{2cm}<{\centering}|}
			\hline
			\multicolumn{1}{|c|}{Parameter}    & Value \\
			\hline
			FOV ${\Psi _{\rm{FOV}}}$& ${60^ \circ }$  \\
			\hline
			Lambertian order $m$ &  $1$  \\
			\hline
			PD's physical receiving area ${A_{\rm PD}}$ & $1{\rm{ cm}^2}$  \\
			\hline
			The gain of the optical
			concentrator ${g_{\rm{f}}}$ & $1$  \\
			\hline
			The area of the indoor scene surface $A_r$
			& $10^ {- 4}{{\rm{m}}^2}$\\
			\hline
		\end{tabular}
	\end{table}

	\begin{figure*}[!h]
		\normalsize	
		\begin{minipage}[b]{0.45\textwidth}
			\centering
			\includegraphics[scale=0.45]{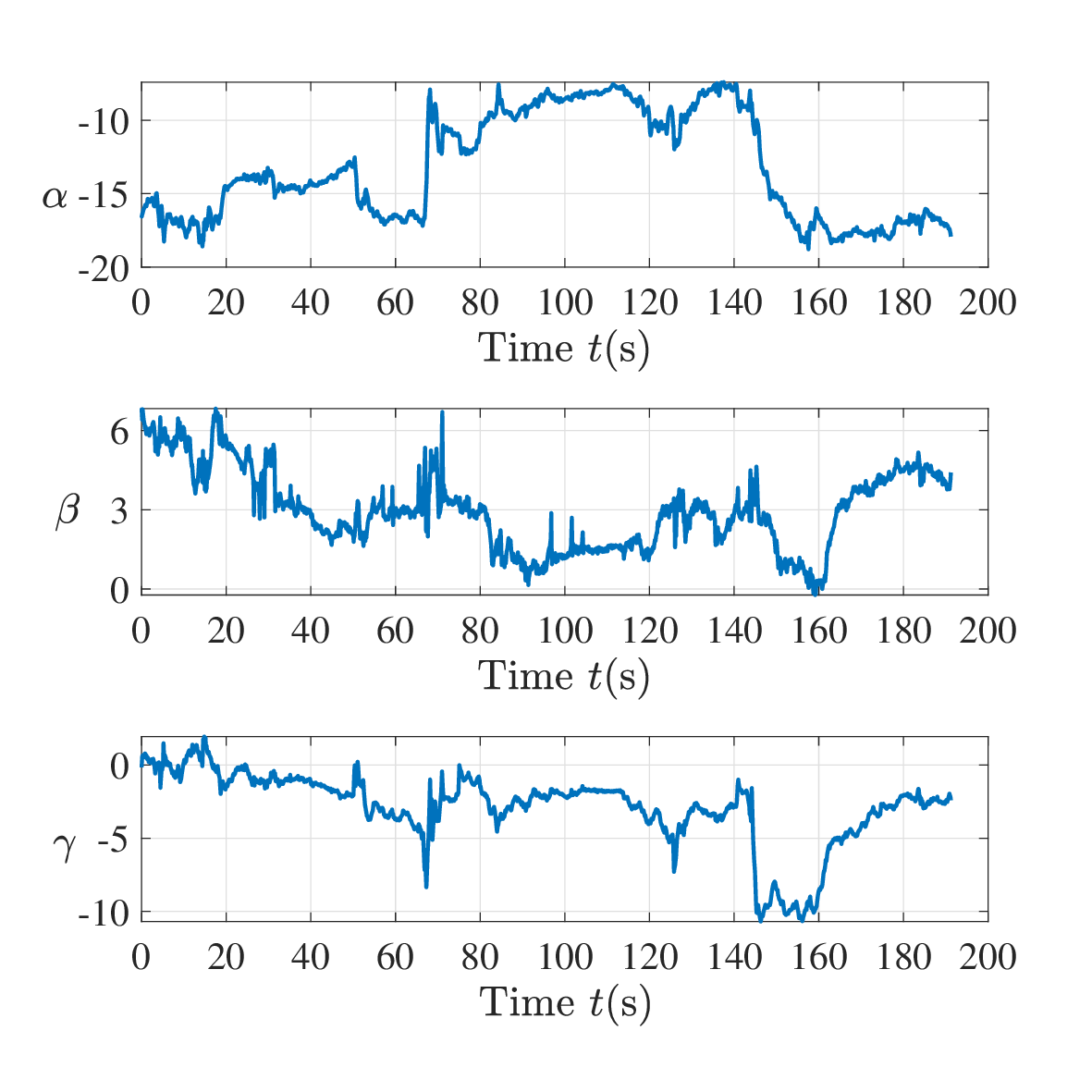}
			\vskip-0.2cm\centering {\footnotesize (a)
			}
		\end{minipage}
		\begin{minipage}[b]{0.45\textwidth}
			\centering
			\includegraphics[scale=0.45]{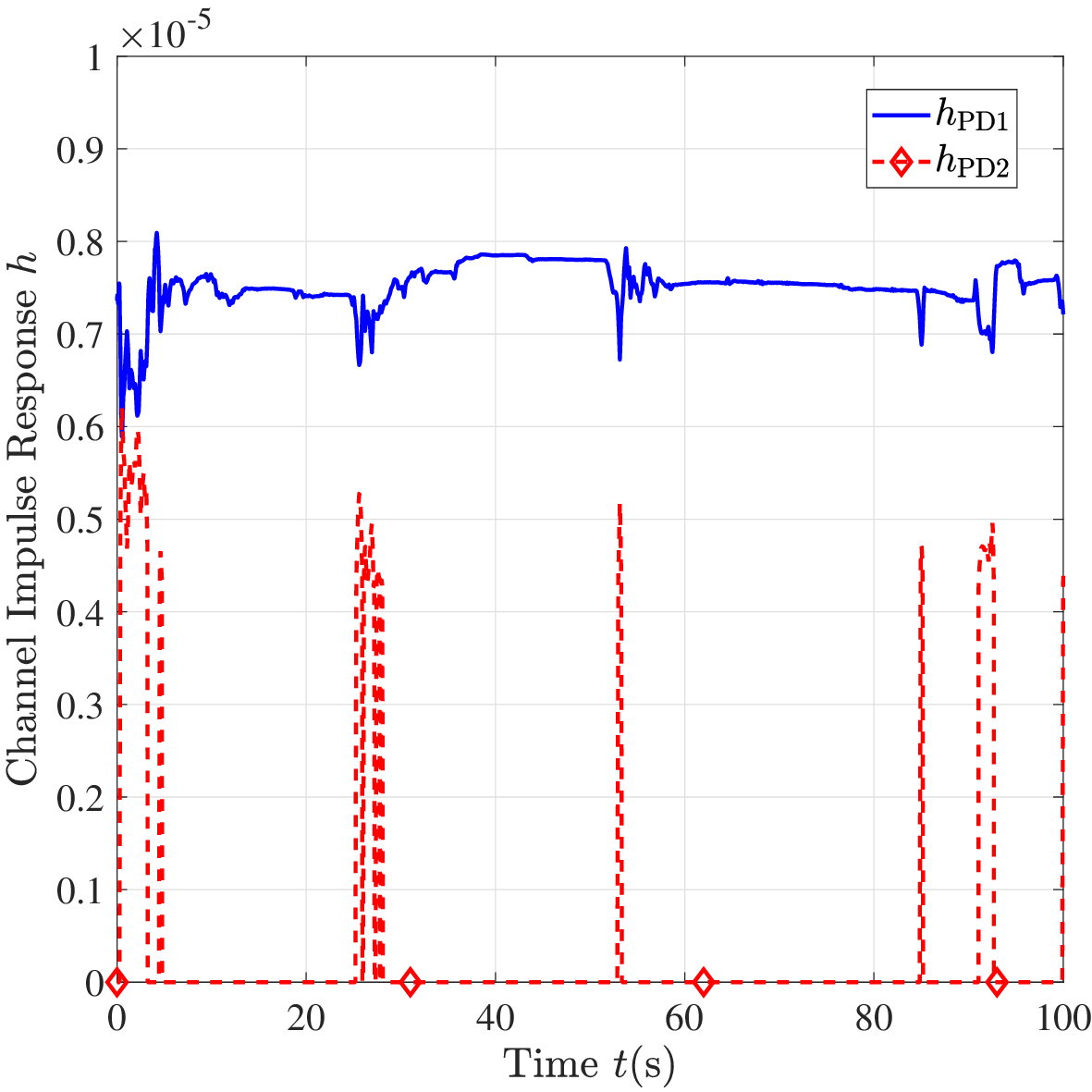}
			\vskip-0.2cm\centering {\footnotesize (b)}
		\end{minipage}\hfill
		\begin{minipage}[b]{0.45\textwidth}
			\centering
			\includegraphics[scale=0.45]{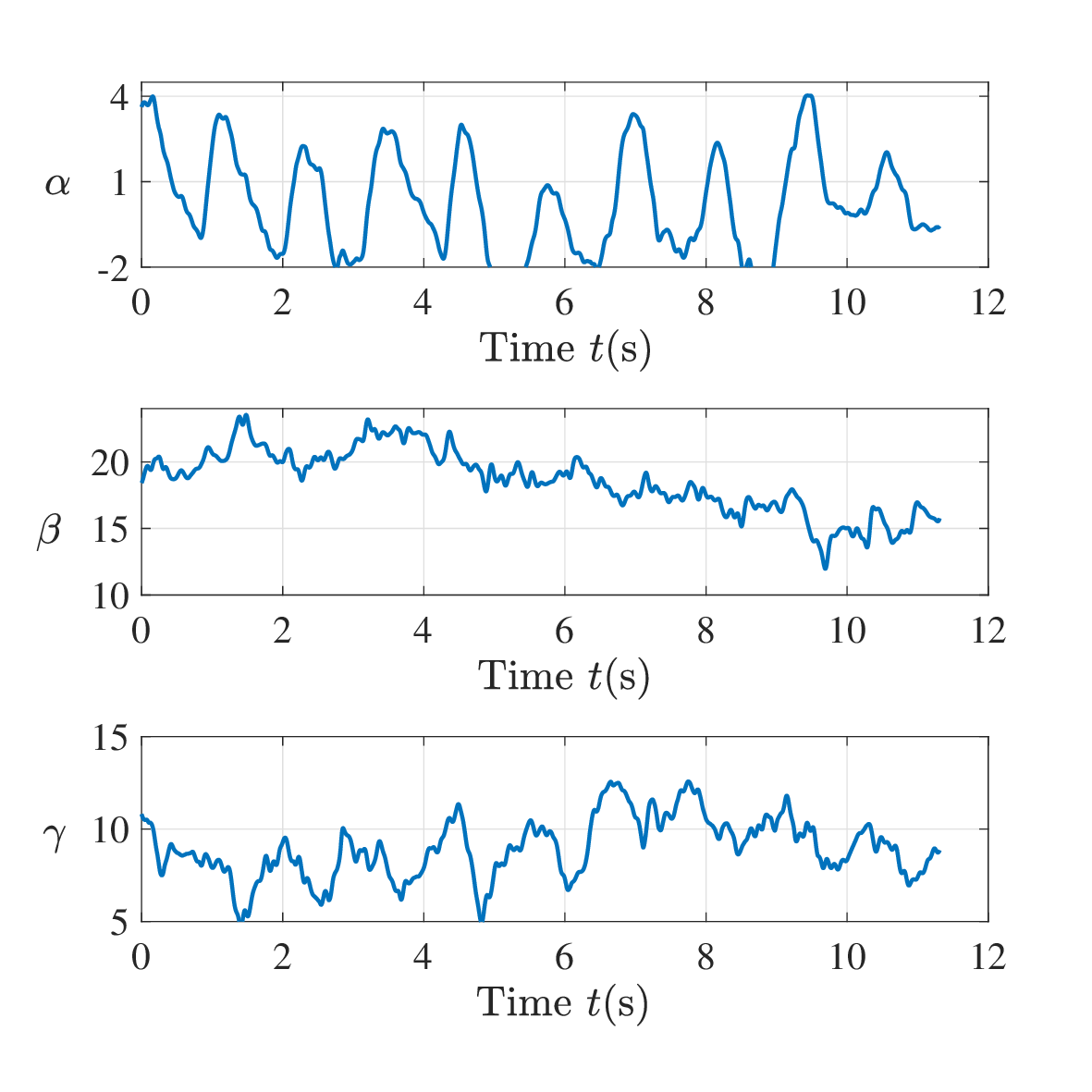}
			\vskip-0.2cm\centering {\footnotesize (c)}
		\end{minipage}\hfill
		\begin{minipage}[b]{0.45\textwidth}
			\centering
			\includegraphics[scale=0.45]{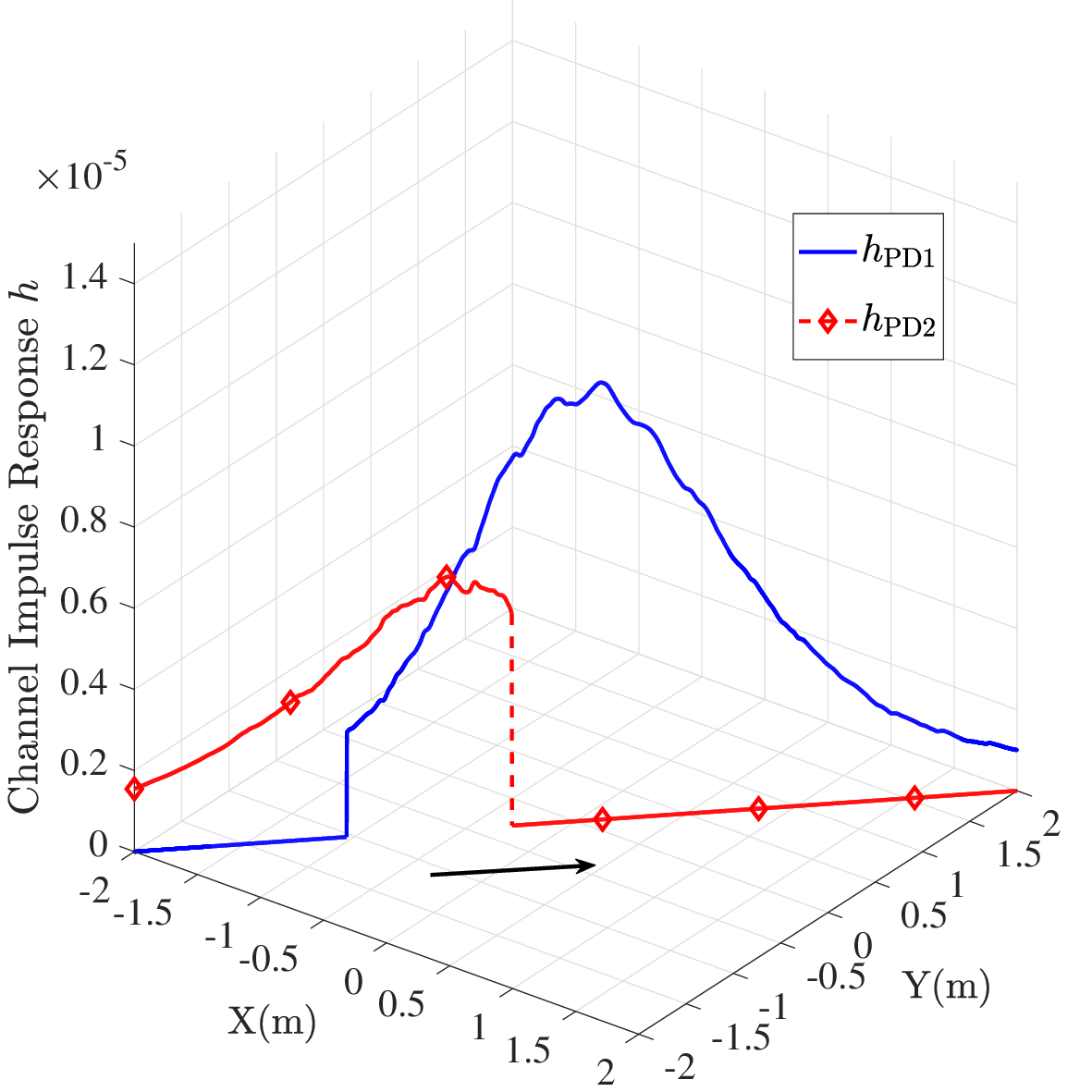}
			\vskip-0.2cm\centering {\footnotesize (d)}
		\end{minipage}\hfill
		\caption{Samples of measurements with different scenarios: (a) angles
			and (b) channel impulse responses in the sitting scenario, (c) angles and (d) channel impulse responses in the walking scenario.}
		\label{h}  
		\vspace*{4pt}
	\end{figure*}
	As shown in Fig. \ref{h}, the rotation angle and channel  of the mobile phone are both measured  in the sitting and walking scenarios. 	
	Fig. \ref{h} (a) and (b) show the angular sequence diagrams and channel impulse responses (CIR) in the sitting scenario where ${{\bf{u}}_l} = {\left[ {0,0,3} \right]^{\rm{T}}}$, ${{\bf{u}}_u} = {\left[ {0,0,1} \right]^{\rm{T}}}$.
	Fig. \ref{h} (c) and (d) show the angular sequence diagrams and CIRs in the walking scenario with ${{\bf{u}}_l} = {\left[ {0,0,3} \right]^{\rm{T}}}$, where testers walk at a speed of $0.6 {\rm{m/s}}$  in the direction of the black arrow. Specifically, the direction of walking can also  be seen in Fig. \ref{h} (d).

	From Fig. \ref{h} (a) and (c), it is observed  that the roll angle $\beta $ is greater than $0$, which means that the tester tends to hold the phone at a tilted angle. Moreover, comparing Fig. \ref{h} (a) with (c), it is observed that the tester's posture of holding the phone is different in various scenarios, which reveals that the real experiment measurement  in the channel modeling is critical.	

	As observed in  Fig. \ref{h} (b), the value of the channel  $h_{\rm{PD}1}$ is higher than that of the channel $h_{\rm{PD}2}$.
	The reason is that the channel $h_{\rm{PD}2}$ is limited by the FOV. The position of PD1 has the advantage to receive the transmitted signal, but the position of PD2 is in outage.

	Furthermore, it can be seen from Fig. \ref{h} (d) that in the walking scenario, the channel  varies as the position where the user walks under the LED.
	The gain peaks of two PDs arrive at different  locations since the geometric positions of two PDs are different.
	
	\subsection{Coherence Time Analysis of Mobile LiFi}
	In our experiments, the channel is measured in both sitting and walking  scenarios. In the sitting scenario, the spacing between  two adjacent testing points is $0.5 {\rm{m}}$.	
	It is worth noting that  the PD may experience interruptions when receiving transmit signals throughout the room, which also leads to ineffective channel information of the samples. In this case, it is meaningless to study the correlation of channel information, thus, the coherence time is defined as $0$.
	
\begin{figure}[htbp]
		\normalsize
		\begin{minipage}[b]{0.45\textwidth}
		\centering
			\includegraphics[scale=0.5]{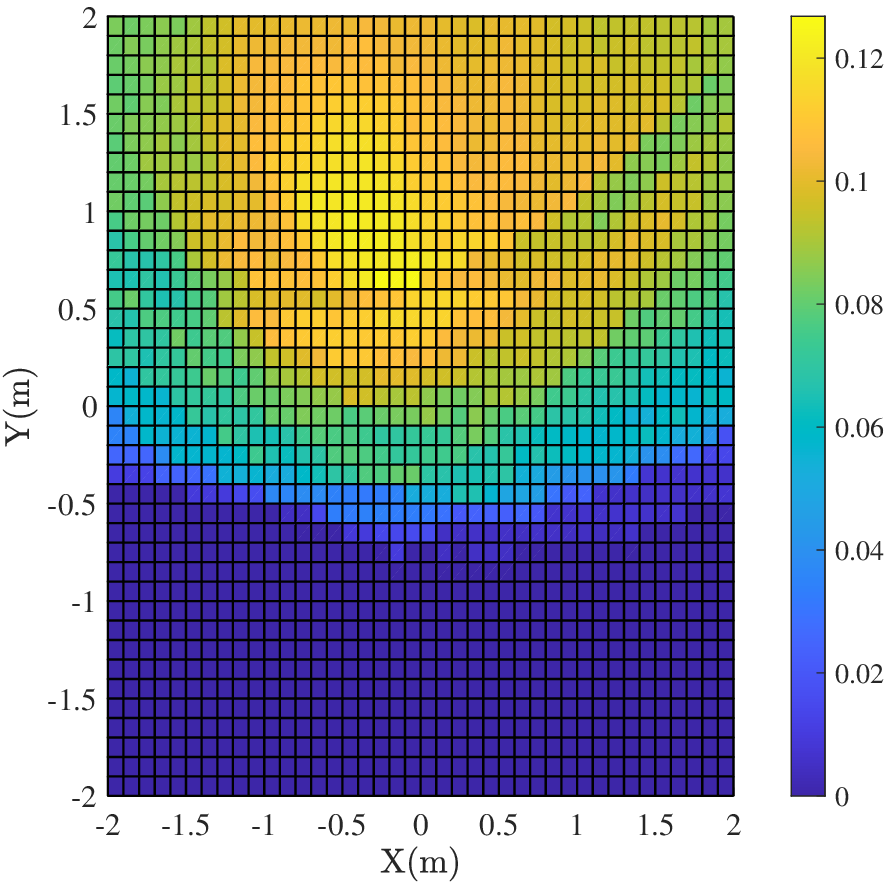}
			\vskip-0.2cm\centering {\footnotesize (a) }
		\end{minipage}
		\begin{minipage}[b]{0.45\textwidth}
		\centering
			\includegraphics[scale=0.5]{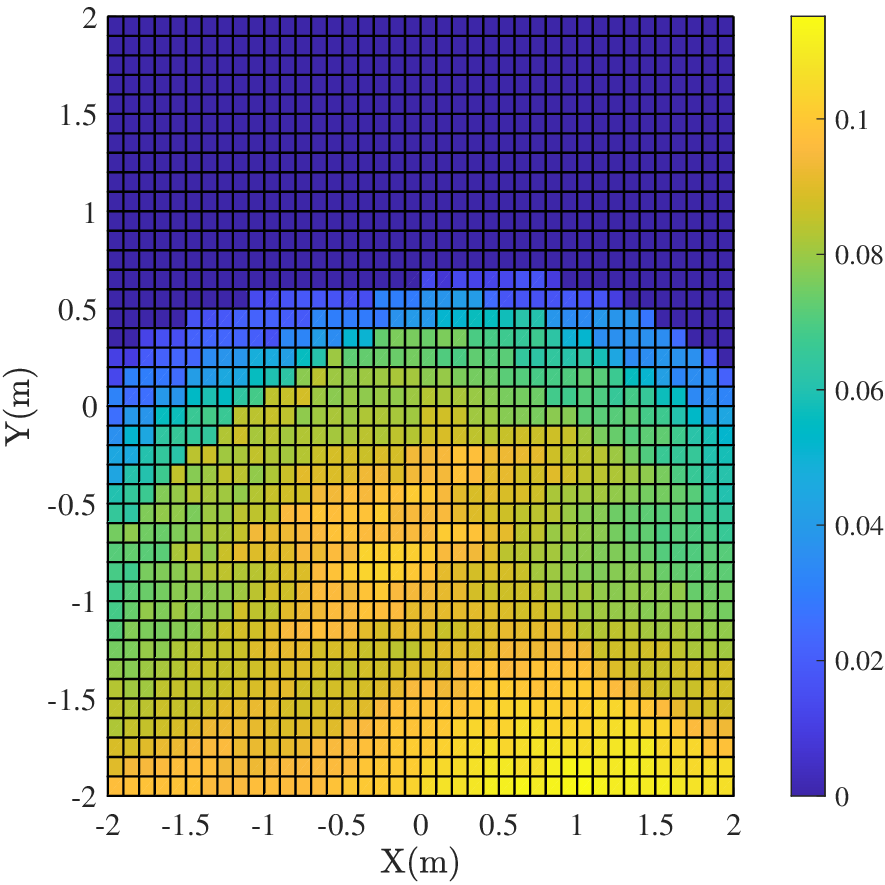}
			\vskip-0.2cm\centering {\footnotesize (b) }
		\end{minipage}\hfill
\begin{minipage}[b]{0.45\textwidth}
	\centering
	\includegraphics[scale=0.35]{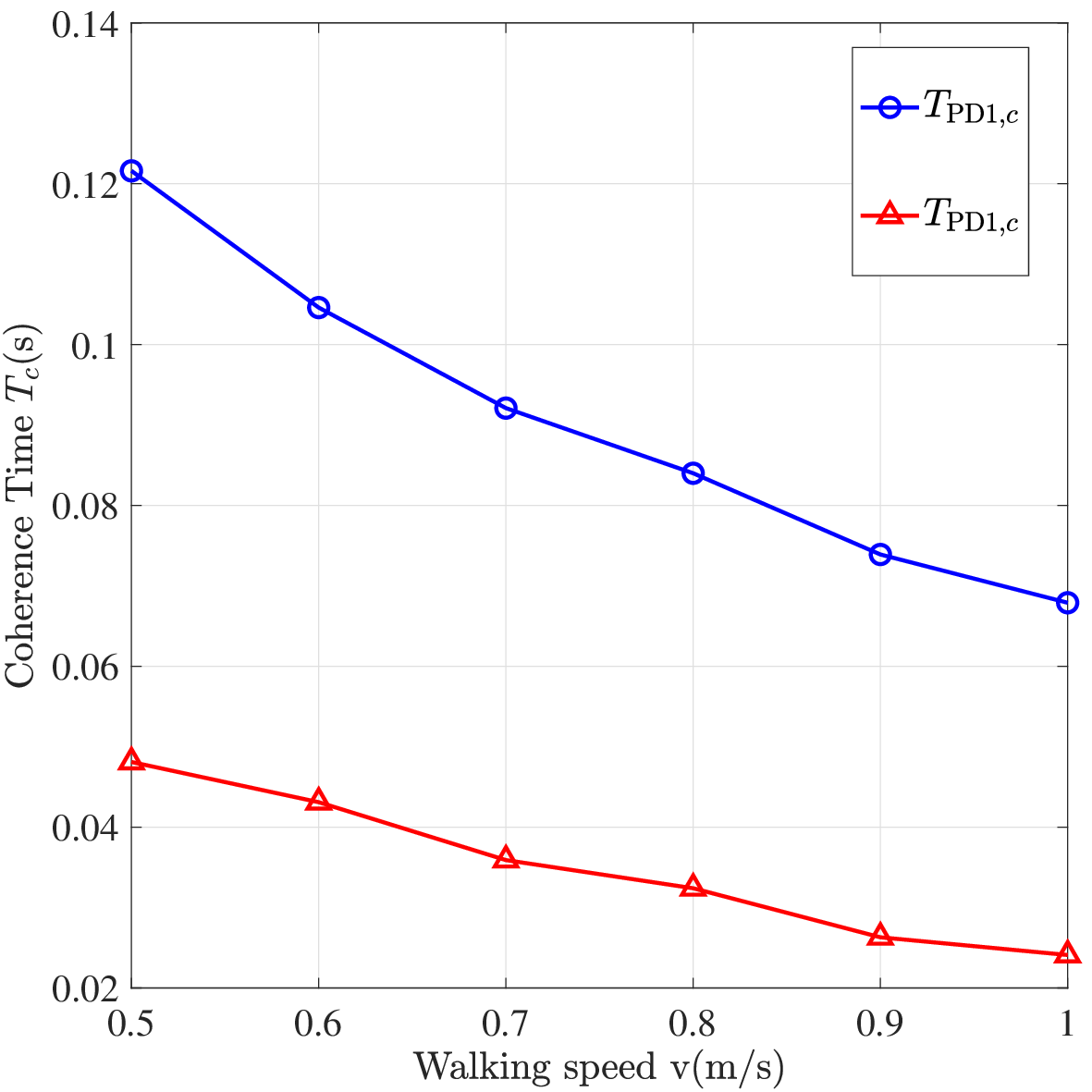}
	\vskip-0.2cm\centering {\footnotesize (c) }
\end{minipage}
		\caption{The coherence time distribution of rooms with ${{\eta _{{\rm{th}}}}}=0.99$: (a) $T_{{\rm{PD}1},c}$ in the sitting scenario, (b) $T_{{\rm{PD}2},c}$ in the sitting scenario, and (c) coherence time distribution in the walking scenario.}
		\label{xiangguanshijian_sit}
		\vspace*{4pt}
	\end{figure}

	Assuming that ${{\eta _{{\rm{th}}}}}=0.99$, ${{\bf{u}}_l} = {\left[ {0,0,3} \right]^{\rm{T}}}$,
	the coherence time $T_{{\rm{PD}1},c}$ of PD1 and $T_{{\rm{PD}2},c}$ of PD2  are calculated at different locations, which are presented in Fig. \ref{xiangguanshijian_sit}. Specifically, Fig. \ref{xiangguanshijian_sit} (a) shows that the channel coherence time  of PD1 is related to the user position, and shows that the coherence time of the user position on the upper Y-axis is higher than that on the lower Y-axis. This is because when the user position is under the lower Y-axis, the PD1 cannot receive the light source (see the PD1 location in Fig. \ref{map}). Thus, the channel is intermittent due to the FOV  restriction, which leads to a lower channel correlation.
	On the contrary, Fig. \ref{xiangguanshijian_sit} (b) shows  that for  PD2, the coherence time of the user position on the upper Y-axis is less than that on the lower Y-axis. The reason is similar to that of  Fig. \ref{xiangguanshijian_sit} (a).
	In walking scenario, as shown in Fig. \ref{xiangguanshijian_sit} (c), the coherence time is different  with various receiving PDs. Furthermore, we can observe that the coherence time decreases with the walking speed.
	
\begin{figure}[htbp]
	\begin{minipage}[b]{0.45\textwidth}
		\centering
		\includegraphics[scale=0.35]{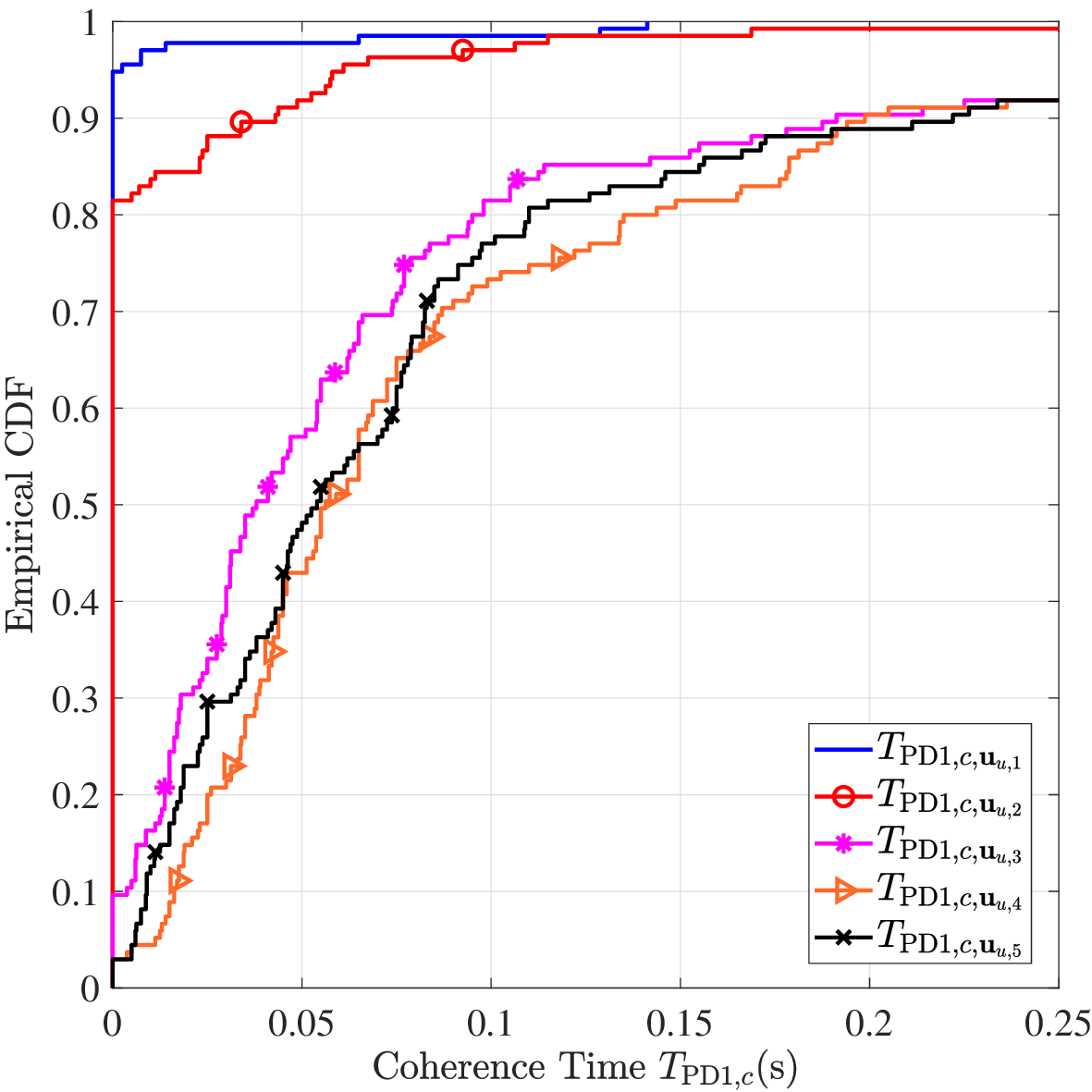}
		\vskip-0.2cm\centering {\footnotesize (a)}
	\end{minipage}\hfill
	\begin{minipage}[b]{0.45\textwidth}
		\centering
		\includegraphics[scale=0.35]{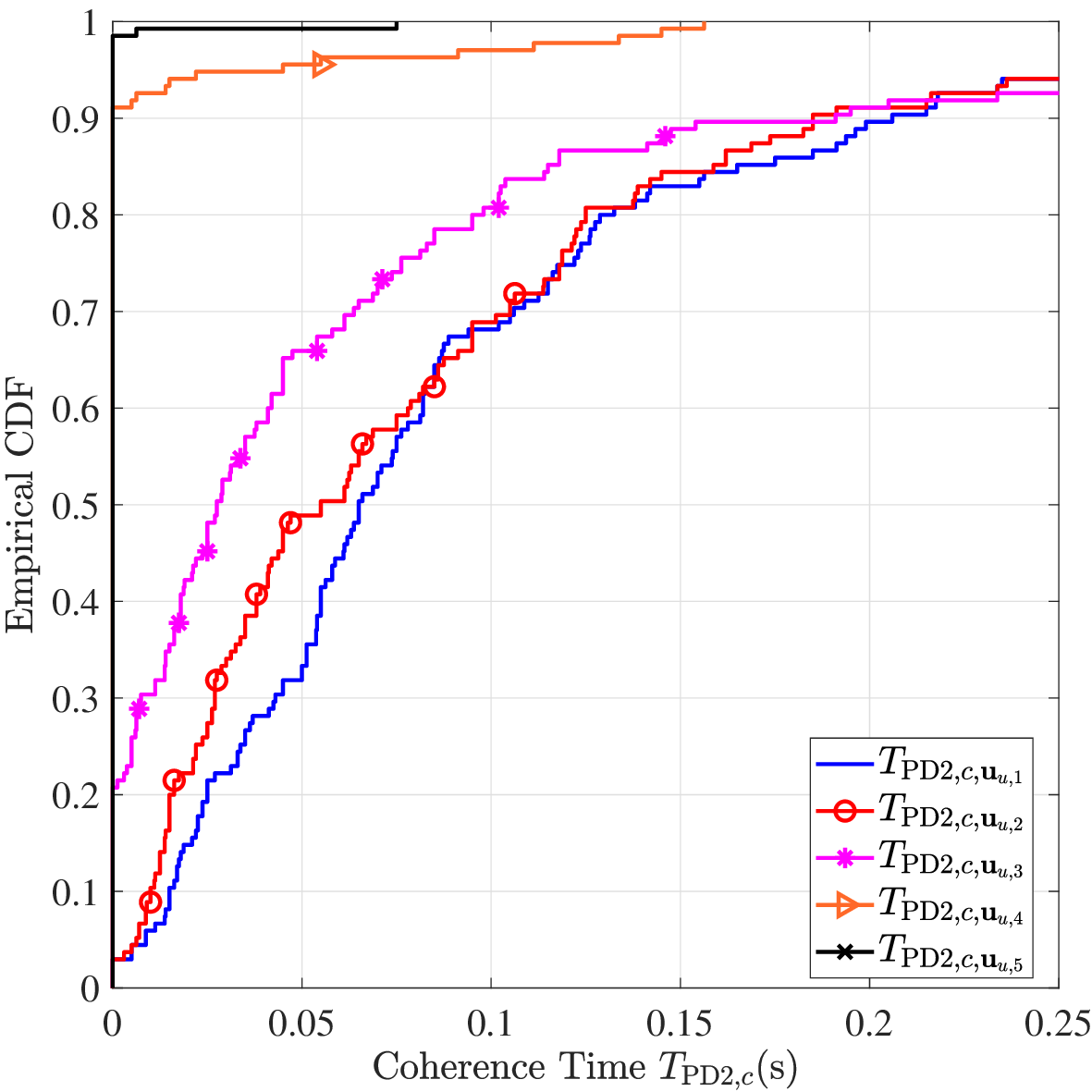}
		\vskip-0.2cm\centering {\footnotesize (b)}
	\end{minipage}\hfill
	\caption{The empirical CDF of the coherence time with the  threshold ${{\eta _{{\rm{th}}}}}=0.99$ in the sitting scenario: (a) $T_{\rm{PD}1,c}$, (b) $T_{\rm{PD}2,c}$.}
	\label{xiangguanshijian_sit_CDF}  
\end{figure}

	Then, we measure and record data at five evenly spaced locations on the diagonal of the room. The empirical cumulative distribution function (CDF) of the coherence time with the  threshold ${{\eta _{{\rm{th}}}}}=0.99$ is shown in Fig. \ref{xiangguanshijian_sit_CDF},
	and the specific coherence time data are given in Table \ref{tableIII}. It can be seen that the coherence time is different at various locations.
	Moreover, the coherence time is influenced by the PD position.

\begin{table}[htbp]
	\centering
	\caption{The coherence time at the  threshold ${{\eta _{{\rm{th}}}}}=0.99$.}	
	\label{tableIII}
	\begin{tabular}{|c|c|c|c|c|c|}
		\hline
		\diagbox {UE Location}{Coherence Time $\left({\rm s}\right)$}  & $T_{{\rm{PD}1},c}$ & $T_{{\rm{PD}2},c}$  \\
		\hline		
		$\begin{array}{*{20}{c}}
			{{{\bf{u}}_{u,1}}}=
			{{{\left[ { - 2, - 2,1} \right]}^{\rm{T}}}}
		\end{array}$ &0.000&	0.065\\
		\hline		
		$\begin{array}{*{20}{c}}
			{{{\bf{u}}_{u,2}}}=
			{{{\left[ { - 1, - 1,1} \right]}^{\rm{T}}}}
		\end{array}$ &0.000&	0.055\\
		\hline
		$\begin{array}{*{20}{c}}
			{{{\bf{u}}_{u,3}}}=
			{{{\left[ { 0, 0,1} \right]}^{\rm{T}}}}
		\end{array}$ &0.038&	0.028\\
		\hline
		$\begin{array}{*{20}{c}}
			{{{\bf{u}}_{u,4}}}=
			{{{\left[ { 1, 1,1} \right]}^{\rm{T}}}}
		\end{array}$ &0.056&	0.000\\
		\hline
		$\begin{array}{*{20}{c}}
			{{{\bf{u}}_{u,5}}}=
			{{{\left[ { 2, 2,1} \right]}^{\rm{T}}}}
		\end{array}$ &0.054&	0.000\\
		\hline
	\end{tabular}
\end{table}


\section{Communication Performance of Mobile LiFi}
In this section, we  investigate the communication rate in SISO, signal-input multiple-output (SIMO), multiple-input signal-output (MISO),  and MIMO  scenarios  to  describe the communication performance of the Mobile LiFi system.

\subsection{Achievable Rate of M-PAM}
In practical communication systems, the transmitted signal is drawn from a discrete signal constellation. The input signal $s$ of the LiFi link is non-negative with $M$ elements. The set of discrete points $\Omega $ is defined as
\begin{align}
	\Omega  \buildrel \Delta \over = \left\{ {s\left| {\begin{array}{*{20}{c}}
				{\Pr \left( {s = {s_k}} \right) = {p_k},0 \le {x_k} \le \hat{A},k = 1, \ldots ,M}\\
				{\sum\limits_{k = 1}^M {{p_k}}  = 1,\sum\limits_{k = 1}^M {{p_k}} {s_k} \le \Phi ,\sum\limits_{k = 1}^M {{p_k}} s_k^2 \le \hat \varepsilon ,{s_k} \in \mathbb{R}}
		\end{array}} \right.} \right\},
\end{align}
where $ {s_k}$ represents m-pulse-amplitude-modulation (M-PAM) discrete constellation points; ${p_k}$ is the probability that $s$ is equal to $ {s_k}$; the parameters $\hat{A}$, $\Phi $, and $\hat \varepsilon $ represent the instantaneous optical power, average optical power, and electrical power threshold of the input signal, respectively.

The signal of the LiFi link is modulated by the M-PAM method, where the bandwidth is denoted as $B$. For single PD and single LED (i.e., signal-input signal-output, SISO) scenario, the received symbol $y$ can be written as
\begin{align}
	y=hs+z,
\end{align}	
where $z$ obeys a Gaussian distribution with a mean of $0$ and a variance of $\sigma^{2}$, i.e., $z \sim$ $\mathcal{N}\left(0, \sigma^{2}\right)$. Thus, the achievable data rate $R_1$ of the SISO scenario can be expressed as
\begin{align}\label{R_i}
	\begin{split}
		{R_1} &= I\left( {s;{y}} \right) \\
		&=  - \frac{B}{{\ln 2}} - 2B\sum\limits_{k = 1}^M {{p_k}} {\mathbb{E}_z}\left\{ {{{\log }_2}\sum\limits_{m = 1}^M {{p_m}} \exp \left( {{\Lambda _{k,m}}} \right)} \right\},
	\end{split}
\end{align}
where ${\Lambda _{k,m}} \buildrel \Delta \over =   - \frac{{{{\left( {{h}\left( {{s_k} - {s_m}} \right) + \sqrt B z} \right)}^2}}}{{2B{\sigma ^2}}}$.

For the single-input multiple-output (SIMO) scenario with $N$ PDs and single LED, the received symbol $y$ can be written as
\begin{align}\label{y_i}
	y = {{\bm{\omega }}^{\rm{T}}}{\bf{h}}s + z,
\end{align}
where $\bm{\omega }=\left[\omega_{1}, \cdots, \omega_{N}\right]^{\mathrm{T}} \in \mathbb{R}^{N \times 1}$ represents the receiving
beamforming for the $N$ PDs, and $\mathbf{h} = \left[h_{1}, \cdots, h_{N}\right]^{\mathrm{T}} \in \mathbb{R}^{N \times 1}$. The achievable data rate $R_2$ of the SIMO scenario can be expressed as
\begin{align}
	{R_2} = & - \frac{B}{{\ln 2}} - 2B\sum\limits_{k = 1}^M {{p_k}{\mathbb{E}_z}\left\{ {{{\log }_2}\sum\limits_{m = 1}^M {{p_m}} } \right.} \nonumber \\
	&\times  \exp \left( { - \frac{{{{\left( {{\bm{\omega}}^{\rm{T}}{{\bf{h}}}\left( {{s_k} - {s_m}} \right) + \sqrt B z} \right)}^2}}}{{2B \sigma ^2}}} \right).
\end{align}		

For the multiple-input signal-output (MISO) scenario with single PD and $K$ LEDs, the received symbol $y$ can be written as
\begin{align}
	y = {{\bf{q}}^{\rm{T}}}{\bf{h}}s + z,
\end{align}
where $\mathbf{q}=\left[q_{1}, \cdots, q_{K}\right]^{\mathrm{T}} \in \mathbb{R}^{K \times 1}$ represents the transmitting beamforming for the $K$ LEDs, and $\mathbf{h}=\left[h_{1}, \cdots, h_{K}\right]^{\mathrm{T}} \in \mathbb{R}^{K \times 1}$. The achievable data rate $R_{3}$ of the MISO case is expressed as
\begin{align}
	{R_{3}} = & - \frac{B}{{\ln 2}} - 2B\sum\limits_{k = 1}^M {{p_k}{\mathbb{E}_{z}}\left\{ {{{\log }_2}\sum\limits_{m = 1}^M {{p_m}} } \right.}\nonumber \\
	&\times   \exp \left( { - \frac{{{{\left( {{\bf{q}}^{\rm{T}}{{\bf{h}}}\left( {{s_k} - {s_m}} \right) + \sqrt B z} \right)}^2}}}{{2B \sigma^2}}} \right).
\end{align}

For the MIMO scenario with $K$ LEDs and $N$ PDs, the received symbol $y$ can be written as
\begin{align}
	{y} = {{\bf{q}}^{\rm{T}}}{\bf{H}}{{\bm{\omega}}}s + z,
\end{align}		
where $\mathbf{q}=\left[q_{1}, \cdots, q_{K}\right]^{\mathrm{T}} \in \mathbb{R}^{K \times 1}$ represents the transmitting
beamforming for the $K$ LEDs, and $\mathbf{H}=\left[{\mathbf{h}_{1}}, \cdots, {\mathbf{h}_{K}}\right]^{\mathrm{T}} \in \mathbb{R}^{K \times N}$. The achievable data  rate $ R_4$ of the MIMO scenario can be expressed as	
\begin{align}
	{ R_4} = & - \frac{B}{{\ln 2}} - 2B\sum\limits_{k = 1}^M {{p_k}{\mathbb{E}_{z}}\left\{ {{{\log }_2}\sum\limits_{m = 1}^M {{p_m}} } \right.}\nonumber \\
	&\times  \exp \left( { - \frac{{{{\left( {{\bf{q}}^T{\bf{H}}{{\bm{\omega}}}\left( {{s_k} - {s_m}} \right) + \sqrt B z} \right)}^2}}}{{2B{\sigma ^2}}}} \right).
\end{align}
\subsection{ Achievable Rate Illustration of Mobile LiFi}
\begin{figure}[!t]
	\begin{minipage}[b]{0.45\textwidth}
		\centering
		\includegraphics[scale=0.35]{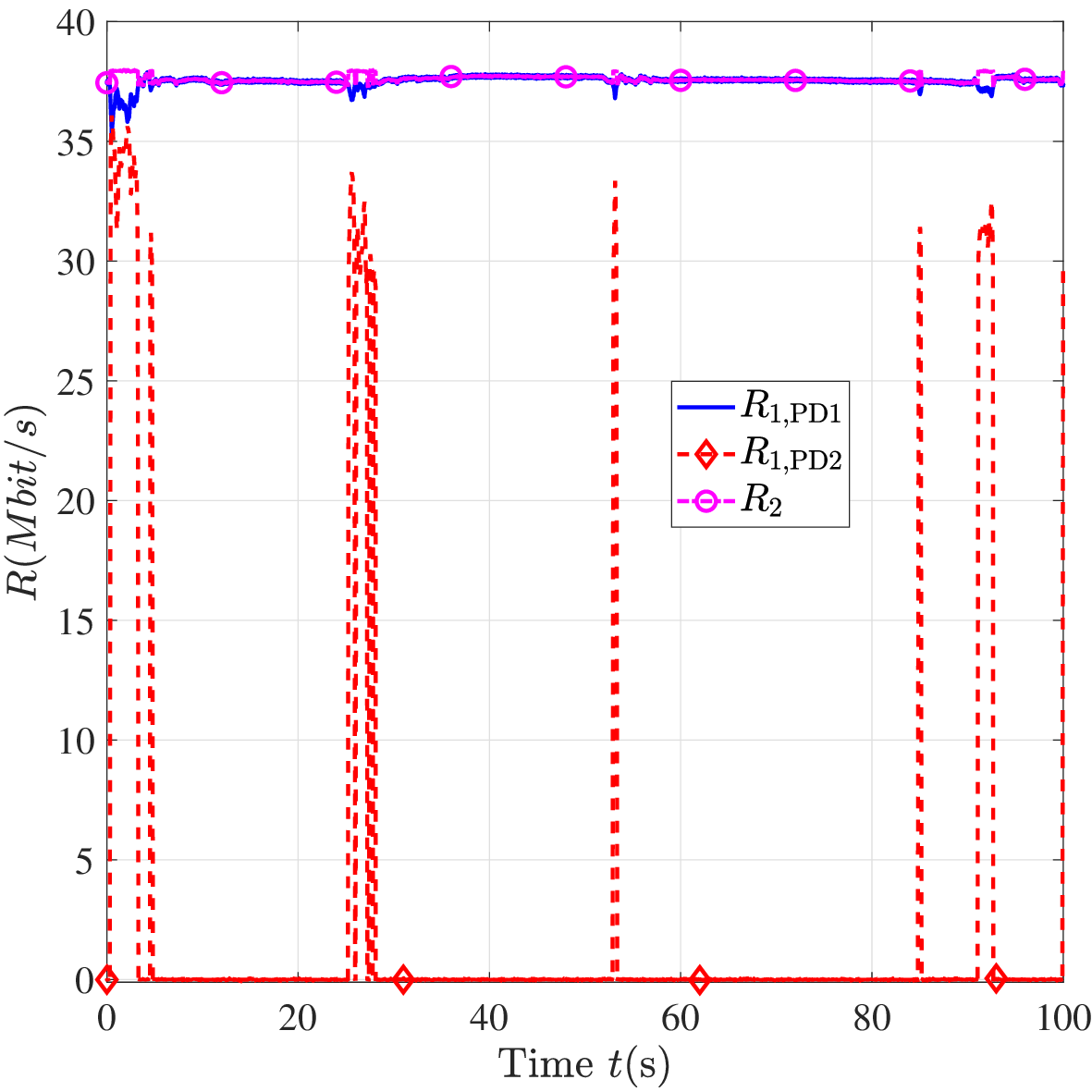}
		\vskip-0.2cm\centering {\footnotesize (a)}
	\end{minipage}
	\begin{minipage}[b]{0.45\textwidth}
		\centering
		\includegraphics[scale=0.35]{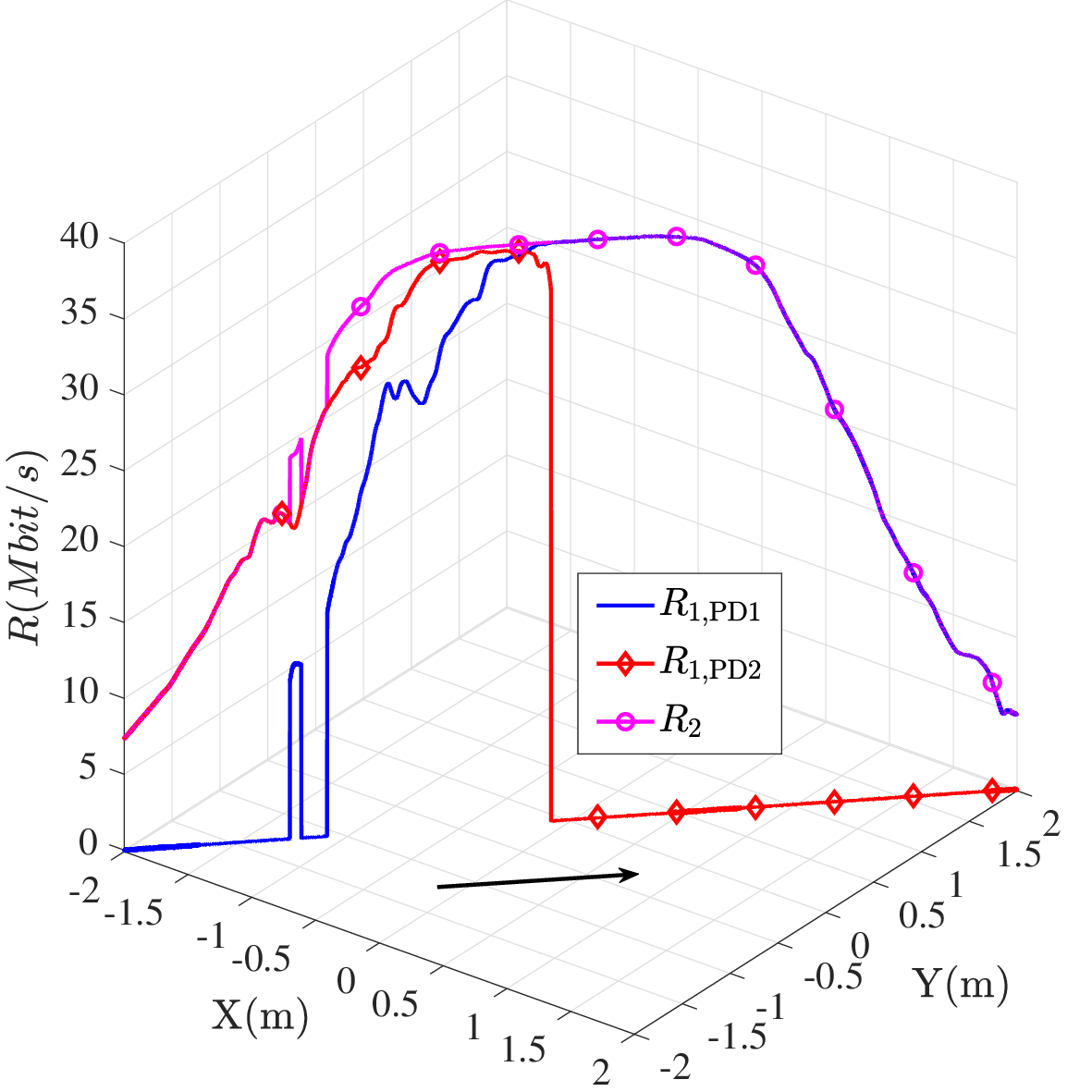}
		\vskip-0.2cm\centering {\footnotesize (b)}
	\end{minipage}\hfill
	\begin{minipage}[b]{0.45\textwidth}
		\centering
		\includegraphics[scale=0.35]{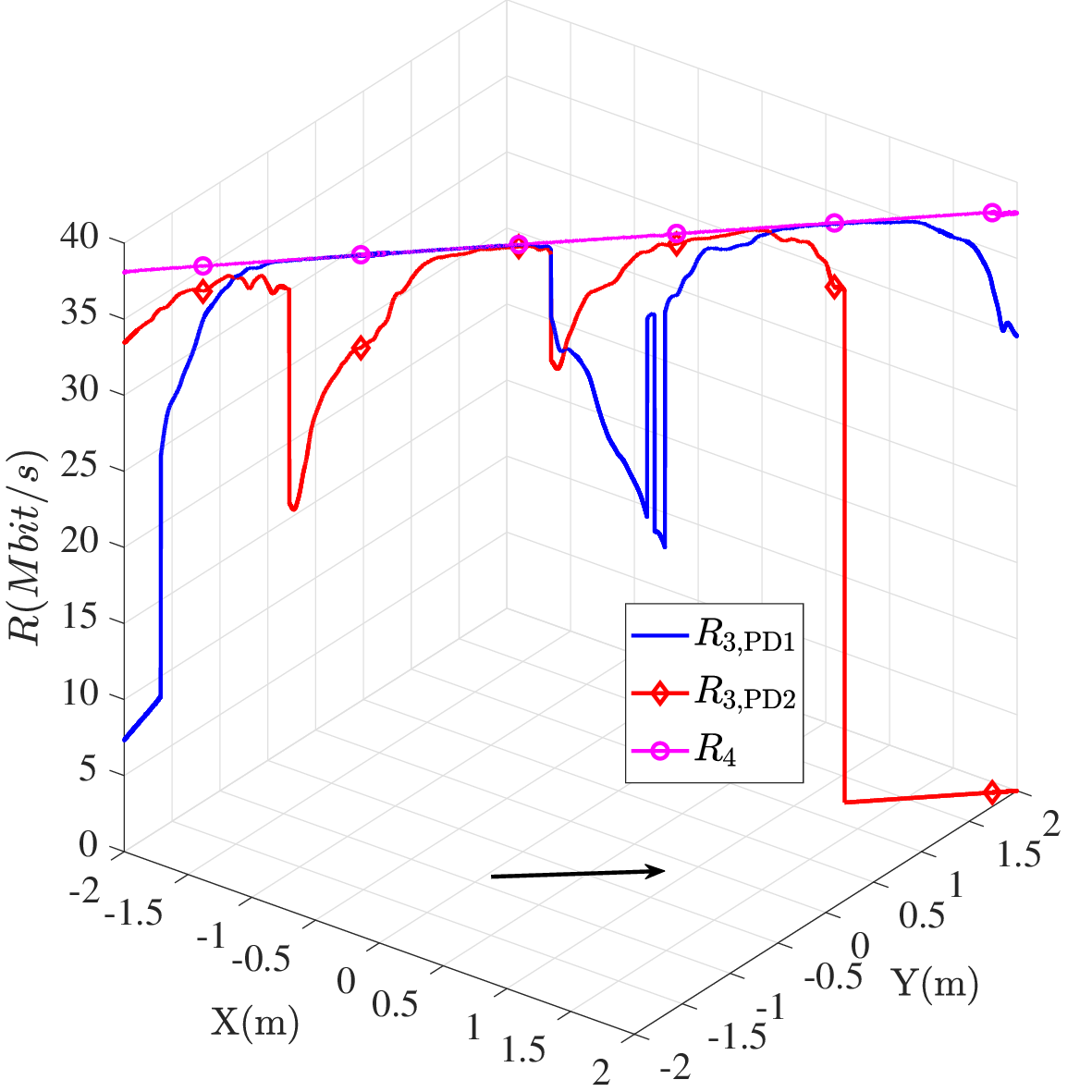}
		\vskip-0.2cm\centering {\footnotesize (c)}
	\end{minipage}\hfill
	
	\caption{Sequence diagram of rate $R(\rm Mbit/s)$ for different scenarios: (a) consider one LED in the sitting scenario, (b) consider one LED in the walking scenario, and (c) consider  multiple LEDs in the walking scenario.}
	\label{R}  
	\vspace*{4pt}
\end{figure}

Assuming 2PAM is adopted and the bandwidth $B = 20{\rm{MHz}}$, Fig. \ref{R} (a) depicts the variation of the rate $R$  with the time $t$ when considering SISO and SIMO of sitting scenarios with one LED, where ${{\bf{u}}_l} = {\left[ {0,0,3} \right]^{\rm{T}}}$ and ${{\bf{u}}_u} = {\left[ {0,0,1} \right]^{\rm{T}}}$.	
It can be seen that in the SISO scenario, PD1 can always receive the light source with the rate $R_{1,{\rm{PD}}1}$ maintained at $36  {\rm{Mbit/s}}$. On the contrary, PD2  intermittently receives the light source. In the SIMO scenario, when PD1 and PD2 are both considered, the LiFi rate $R_{2}$ is maintained at $37.5 {\rm{Mbit/s}}$, which can ensure stable communication transmission.

In the walking scenario with a speed of  $0.6 {\rm{m/s}}$, Fig. \ref{R} (b) depicts the variation of the rate $R$  with  terminal positions with one LED.
It can be seen that as the tester walks, the PD may not receive the light source with the change of positions when individually considering PD1 and PD2 in the SISO scenario. However, it is more stable to support mobile LiFi when two PDs are considered at the same time.

Moreover,  we deploy five LEDs in a line with the spacing of $1.41{\rm m}$ for the walking scenario. The corresponding rate curve is shown in Fig. \ref{R} (c). Comparing Fig. \ref{R} (b) and (c), we can see that the performance of that PDs receiving the light source is improved by adding multiple LEDs. Moreover, the results show that the rate is stable at $36 {\rm{Mbit/s}}$ for the MIMO scenario.

\section{Channel Estimation of Moblie LiFi}		
Accurate channel estimation is crucial for the performance of a coherent wireless communication system.
In this section, we propose two channel estimation schemes of mobile LiFi, i.e., channel  estimation coding and convolutional neural network based deep residual network (CDRN), to improve the rate of LiFi.

We assume that the channel  is slowly fading and is constant over a short period of time. During this period, we use ${\tilde h}$ to represent the channel gain.
The received signal in the $n$th time slot can be expressed as
\begin{align}\label{y_n}
	y\left[ n \right] = \tilde hs\left[ n \right] + z\left[ n \right],
\end{align}
where  $s\left[ n \right]$ represents input signal, $z\left[ n \right]$ represents the independent and identically distributed additive white Gaussian noise with zero mean and the variance of ${\sigma ^2}$.

\subsection{Channel  Estimation Coding of Mobile LiFi}
In a mobile LiFi system , the maximum optical power and average optical power limits for sending information must satisfy
\begin{align}
	&0 \le s\left[ n \right] \le \hat{\rho}, \\
	&\frac{1}{L}\sum\limits_{n = 1}^L {s\left[ n \right] = \hat{\Phi} },
\end{align}
where $\hat{\rho} $ represents the maximum optical power and $\hat{\Phi} $ represents the average transmitted optical power.

To estimate ${\tilde h}$, we use a linear decoding estimator $w\left[ n \right]$  to obtain the estimate $\hat h$, given by
\begin{align}\label{coding_hath}
	\hat h = \sum\limits_{n = 1}^L {w\left[ n \right]y\left[ n \right]} .
\end{align}
By using the zero-forcing decoder, we have
\begin{align}\label{zf}
	\sum\limits_{n = 1}^L {\left( {w\left[ n \right]s\left[ n \right]} \right)}  = 1.
\end{align}

{Therefore, the estimated channel $\hat h $ can be expressed as}
\begin{align}\label{hat_h}
	\begin{split}
		\hat h
		= {\tilde h} + \sum\limits_{n = 1}^L {w\left[ n \right]} z\left[ n \right].
	\end{split}
\end{align}

{  Based on \eqref{hat_h}, $\hat{h}$ is related to a zero-mean Gaussian noise with variance of ${\sigma ^2}\sum\limits_{n = 1}^L {{w^2}\left[ n \right]} $.}

Let ${\bf{w}} \buildrel \Delta \over = {\left[ {\begin{array}{*{20}{c}}
			{{w}\left[ 1 \right]}& \cdots &{{w}\left[ L \right]}
	\end{array}} \right]^T}$ and ${\bf{a}} = \left[ {s\left[ 1 \right], \ldots ,s\left[ n \right]} \right]
$, the goal is to design the pilot signal pattern ${\bf{a}}$ and ${\bf{w}}$ to minimize total noise variance.
According to (\ref{zf}), ${\bf{aw}} = 1$ and $	{{\bf{w}}^T}{{\bf{w}}} $ can be written as
\begin{align}
	\begin{split}
		{{\bf{w}}^T}{\bf{w}}
		= {\left( {{\bf{w}} - {{\bf{a}}^\dag }} \right)^T}\left( {{\bf{w}} - {{\bf{a}}^\dag }} \right) + {\left( {{{\bf{a}}^\dag }} \right)^T}{{\bf{a}}^\dag }.
	\end{split}
\end{align}

For any given $\bf a$ at the transmitters, the optimal $\bf w$ at the receiver is  ${\bf{a}}^\dag$, i.e. ${\bf w}={{\bf{a}}^\dag}$. Hence, the optimization problem is simplified to design the optimal ${\bf{a}}$ to minimize the total noise variance, i.e., minimize
\begin{align}
	\begin{split}
		{\rm{Tr}}\left( {{{\left( {{{\bf{a}}^\dag }} \right)}^T}{{\bf{a}}^\dag }} \right)
		= \frac{1}{{{{\left\| {\bf{a}} \right\|}^2}}}.
	\end{split}
\end{align}

Under the constraints of maximum optical power and average power, the problem of minimizing total noise variance can be modeled as
\begin{subequations}\label{yueshu}
	\begin{align}
		\mathop {{\rm{min}}}\limits_{\bf{a}} &{\rm{ Tr}}\left( {{{\left( {{\bf{a}}{{\bf{a}}^T}} \right)}^{ - 1}}} \right)\\
		s.t.&0 \le s\left[ n \right] \le \hat{\rho} ,\\
		&{\rm{       }}\frac{1}{L}\sum\limits_{n = 1}^L {s\left[ n \right] = \hat{\Phi} }.
	\end{align}
\end{subequations}

Problem \eqref{yueshu} is a convex problem, which can be optimally solved by standard convex optimization solvers such as CVX. Finally, $\hat{h}$ is computed using \eqref{coding_hath}.

\subsection{Channel Estimation of Mobile LiFi Based on CDRN}
To further improve the accuracy of the channel estimation results, a convolutional neural network (CNN)-based deep residual network  is proposed, which implicitly learns the residual noise
from the noisy observations for recovering the channel coefficients.
%
%

\begin{figure}[htbp]
	\centering
	\includegraphics[width=9cm]{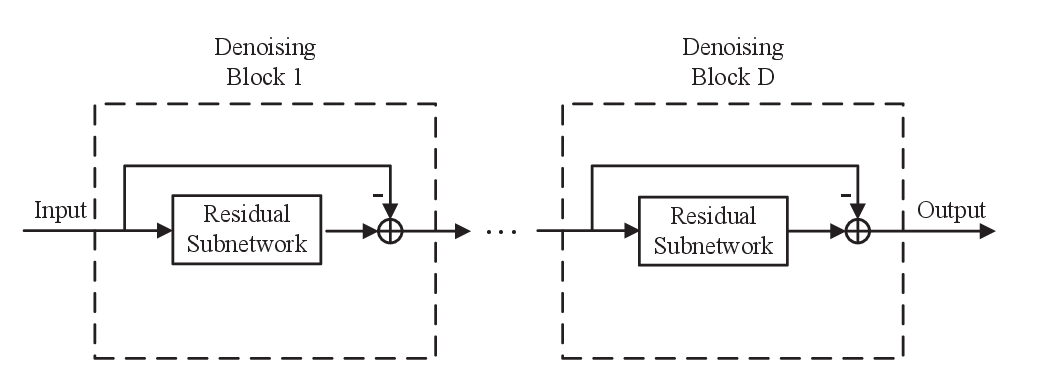}
	\caption{The channel estimation process of mobile LiFi based on  CDRN}
	\label{CDRN1}
\end{figure}
\subsubsection{Channel Estimation Procedure}
The CDRN architecture consists of one input layer, $D$
denoising blocks, and one output layer.
Specifically, each denoising block is cascaded sequentially to gradually enhance the denoising result. Fig. \ref{CDRN1} outlines the channel estimation process to simplify the discussion.

The LS estimator has been widely adopted in practice for its low complexity since it does not require  a prior knowledge of data.
Therefore, we can use the  LS estimator output as the coarse estimated value.  Moreover, based on the channel correlation, the training set can be expressed as
\begin{align}
	{\bf{P}}=\left( {{\bf{\hat h}}_{k}^{LS},{\bf{\hat h}}_{k + 1}^{LS}} \right),
\end{align}
where ${\bf{\hat h}}_{k}^{LS} \buildrel \Delta \over = \left[ {\hat h_{k,1}^{LS}, \cdots ,\hat h_{k,L}^{LS}} \right]$, $\hat h_{k,L}^{LS}$ represents the rough estimate of the channel information in the $k$th coherence time period.
The preliminary estimated channel information is the input of the CDRN, and the estimated channel information is the output of the mapping relationship, which can be formulated as
\begin{align}
	\left\{ {{\bf{\hat h}}_{k}^{},{\bf{\hat h}}_{k + 1}^{}} \right\} = {f_{\vartheta} }\left( {{\bf{\hat h}}_{k}^{LS},{\bf{\hat h}}_{k + 1}^{LS};\vartheta} \right),
\end{align}
where $\vartheta $ represents the parameter set.

\subsubsection{Training}
There are $D$ identical denoising blocks in the CDRN, which are adopted to gradually enhance the denoising performance. Each denoising block consists of a residuals subnetwork and an element subtraction. In a residual subnetwork, each layer uses a combination of "Conv+BN+ReLU" operations. Specifically, the convolution (Conv) and the rectified linear unit  (ReLU) are adopted jointly to explore the spatial features of channel matrices and the batch normalization (BN) is added between them to improve the network stability and the network training speed.
Moreover, a Conv operation is used to extract features in the last layer of the subnetwork to construct the noise matrix. Using the additive principle of noise, the elemental subtraction method is used to remove noise. Table \ref{CDRN} lists the detailed architecture of the CDRN denoising block.
\begin{table}[H]
	\centering
	\caption{The detailed architecture of the CDRN.}	
	\begin{tabular}{ccc}
		\cline{1-3}
		\multicolumn{3}{l}{{\bf{Input:}} Channel matrix ${{\bf{P}}} $}  \\
		\cline{1-3}
		\multicolumn{3}{l}{\bf{Denoising Block: }}    \\
		{\bf{Layers}} & {\bf{Operations}} & {\bf{Filter size}}  \\
		1 & Conv $+\mathrm{BN}+\operatorname{ReLU}$ & $64 \times(3 \times 3 \times 1)$   \\
		$2 \sim 15$ & Conv $+\mathrm{BN}+\operatorname{ReLU}$ & $64 \times(3 \times 3 \times 1)$    \\
		16 & Conv & $1 \times(3 \times 3 \times 64)$   \\
		\cline{1-3}
		\multicolumn{3}{l}{{\bf{Output:}} Channel matrix ${f_{\vartheta}} \left( {\bf{P}} \right)$}   \\
		\cline{1-3}
	\end{tabular}
	\label{CDRN}
\end{table}
For sequentially cascaded denoising blocks, let ${f_d}\left(  \cdot  \right), d \in \left\{ {1, \cdots D} \right\}$ represent the function expression of the $d$th denoising block. Then, the $d$th denoising block is represented as
\begin{align}
	{{\bf{P}}_d} = {{\bf{P}}_{d - 1}} - {f_d}\left( {{{\bf{P}}_{d - 1}}} \right),\forall d,
\end{align}
where ${{\bf{P}}_d}$ and ${{\bf{P}}_{d - 1}}$ represent the input and output of the $d$th denoising block, respectively.
The network  output can be expressed as
\begin{align}
	{f_{\vartheta} }\left( {\bf{P}} \right) = {\bf{P}} - \sum\limits_{d = 1}^D {{f_d}\left( {{{\bf{P}}_{d - 1}}} \right)},
\end{align}
where $\sum\limits_{d = 1}^D {{f_d}\left( {{{\bf{P}}_{d - 1}}} \right)} $ represents the residual noise component. Thus, the obtained channel is the result of denoising.

The total loss function ${J_{Loss}}$ of the network is the normalized mean squared error (NMSE) between the estimated and actual channel responses, which can be calculated as
\begin{align}
	{J_{Loss}} = \frac{{{{\left\| {{\bf{\hat h}}_{k + 1}^{} - {\bf{h}}_{k + 1}^{}} \right\|}^2}}}{{{{\left\| {{\bf{h}}_{k + 1}^{}} \right\|}^2}}}.
\end{align}

\subsection{Methods Evaluation}	
\begin{figure}[!t]
	\centering
	\includegraphics[scale=0.35]{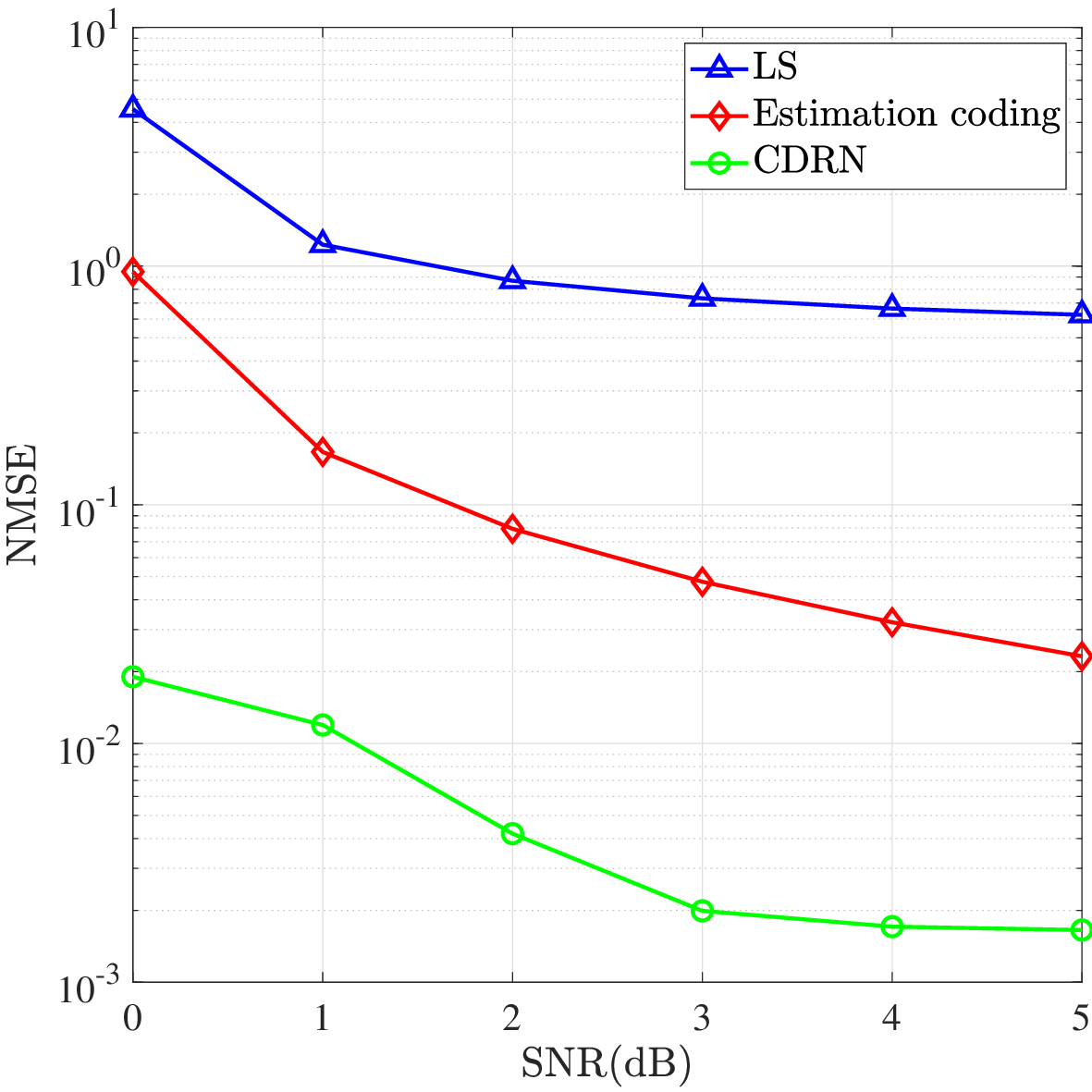}
	\vskip-0.2cm\centering {\footnotesize}
	\caption{Channel estimation  performance among different schemes.}
	\label{Channel estimation}
\end{figure}

NMSE is selected
as the performance metric for  channel estimation.	
Based on the coherence time, we present the NMSE measured for the results of the CDRN, estimation coding, and LS schemes. PD1 is chosen in the sitting scenario to collect the experimental data for the channel estimation.  Assuming ${{\bf{u}}_l} = {\left[ {0,0,3} \right]^{\rm{T}}}$, ${{\bf{u}}_u} = {\left[ {0,0,1} \right]^{\rm{T}}}$, according to the coherence time of TABLE III,  the coherence time is $0.038{\rm s}$.

Fig. \ref{Channel estimation} compares the proposed channel estimation coding, CDRN, and LS schemes.
It can be seen that the channel error decreases as the SNR increases.	 Moreover,
the channel  estimation coding scheme and the proposed CDRN scheme are superior to the LS.  In addition, the proposed CDRN scheme is superior to the channel estimation coding scheme. The reason is that the CDRN intelligently exploits the non-linear spatial features of channels in a data driven approach.

\section{Channel Tracking of Moblie LiFi }

In  practical time-varying systems, the system performance is affected by  channel aging. To address this issue, we propose a neural network-based channel tracking method to estimate more accurate channel information. The neural network can be used to track current CSI in a time-varying scenario.  The channel tracking problem can be viewed as a time sequence prediction problem, which can be solved by the LSTM model. 

\subsection{Channel Tracking Model of Mobile LiFi Based on  LSTM }
\begin{figure}[ht]
	\centering
	\includegraphics[width=9cm]{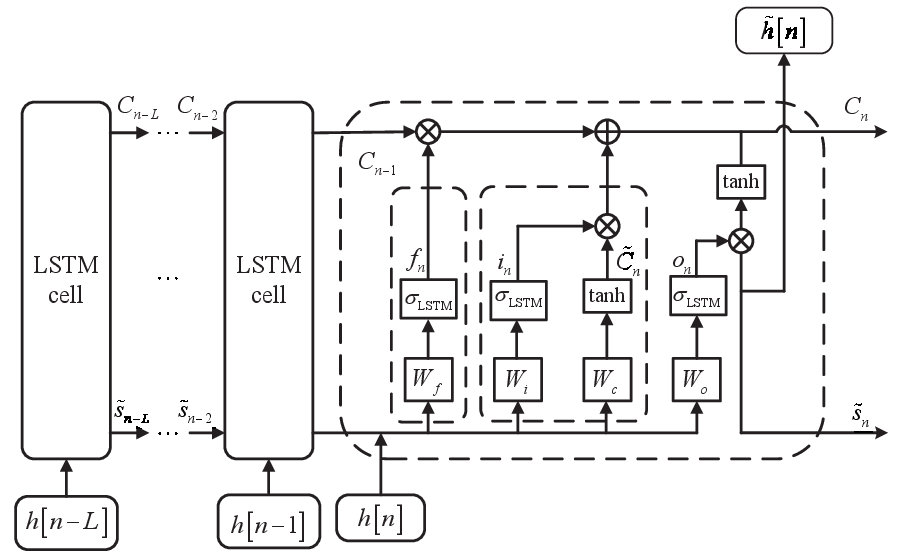}
	\caption{ The channel tracking  structure of mobile LiFi based on LSTM. }
	\label{lstm_system} 
\end{figure}

As shown in Fig. \ref{lstm_system}, each cell of the LSTM model has a complex recursive structure.
The keys to the LSTM model are the cell state and the three structures of gates.
Through the selection of the gates, the LSTM removes or adds channel information of the cell state to control the cell state.
At the $n$th time slot, the gate ${i}_{n}$ in Fig. \ref{lstm_system} is called the input gate, which determines whether the cell adds the current channel information to the cell state ${{C}}_{n}$.
The forget gate ${f}_{n}$ determines whether the cell retains or discards the  previous cell state ${{C}}_{n-1}$, which depends on the current channel information $ h\left[ n \right]$ and the previous hidden state $ {{\tilde{s}}}_{n-1}$. The output gate ${o}_{n}$ determines whether the cell outputs the state $ {{\tilde{s}}}_{n}$ and the prediction  channel state $\tilde h\left[ n \right]$.

 The calculation of an LSTM cell is as follows
\begin{subequations}
	\begin{align}
		&{f}_{n}={\sigma_{\rm{LSTM}}}\left({W}_{f} {h\left[ n \right]}+{W}_{f} {{\tilde{s}}}_{n-1}+{b}_{f}\right),\\
		&\tilde{{C}}_{n}={\phi_{\rm{tanh}}}\left({W}_{c} {h\left[ n \right]}+{W}_{c} {{\tilde{s}}}_{n-1}+{b}_{c}\right),\\
		&{i}_{n}={\sigma_{\rm{LSTM}}}\left({W}_{i } {h\left[ n \right]}+{W}_{i} {{\tilde{s}}}_{n-1}+{b}_{i}\right),\\
		&{C}_{n}=\tilde{{C}}_{n} \otimes {i}_{n}+{C}_{n-1} \otimes {f}_{b},\\
		&{o}_{n}={\sigma_{\rm{LSTM}}}\left({W}_{o} {h\left[ n \right]}+{W}_{o} {{\tilde{s}}}_{n-1}+{b}_{o}\right),\\
		&{{\tilde{h}}\left[ n  \right]}=\tanh \left({C}_{n}\right) \otimes {o}_{n},
	\end{align}
\end{subequations}
where $\tilde{{C}}_{n}$ is an intermediate vector for the cell state; $\sigma_{\rm{LSTM}}(\cdot)$ is the sigmoid function, $\phi_{\rm{tanh}}(\cdot)$ is the hyperbolic
tangent function; ${W}_{f}$ ,
${W}_{c}$, ${W}_{i}$, and $ {W}_{o}$ are the weights of the neural network; ${b}_{f}$ ,
${b}_{c}$, ${b}_{i}$, and ${b}_{o}$ are the biases of the neural network. These
parameters determined by the adequate training control the mapping of the input variables
for every gate.

\subsection{Online Training}
Based on real channels, the online training network does not depend on the entire training dataset and can dynamically adapt to new environments.
In the initial phase, the network is trained based on a small dataset from the real channel.
Then, we normalize the sample data to reduce the influence of different ranges of data and to improve the model convergence speed.
Moreover, the dataset is divided into the training set and test set. The training set is used to train the model and the test set is used for prediction.
Finally, the predicted values are denormalized.

In addition, a normalized channel error $\Delta h$ is selected
as the performance metric of  channel estimation, defined as
\begin{align}
	\Delta {h}\left[ n \right] = \frac{{\left| {{h}\left[ n \right] - {{\hat h}}\left[ n \right]} \right|}}{{{h}\left[ n \right]}}.
\end{align}

\subsection{Channel Tracking Illustration of Mobile LiFi}
\begin{figure}[htbp]
	\centering
	\begin{minipage}[b]{0.45\textwidth}
		\centering
		\includegraphics[scale=0.35]{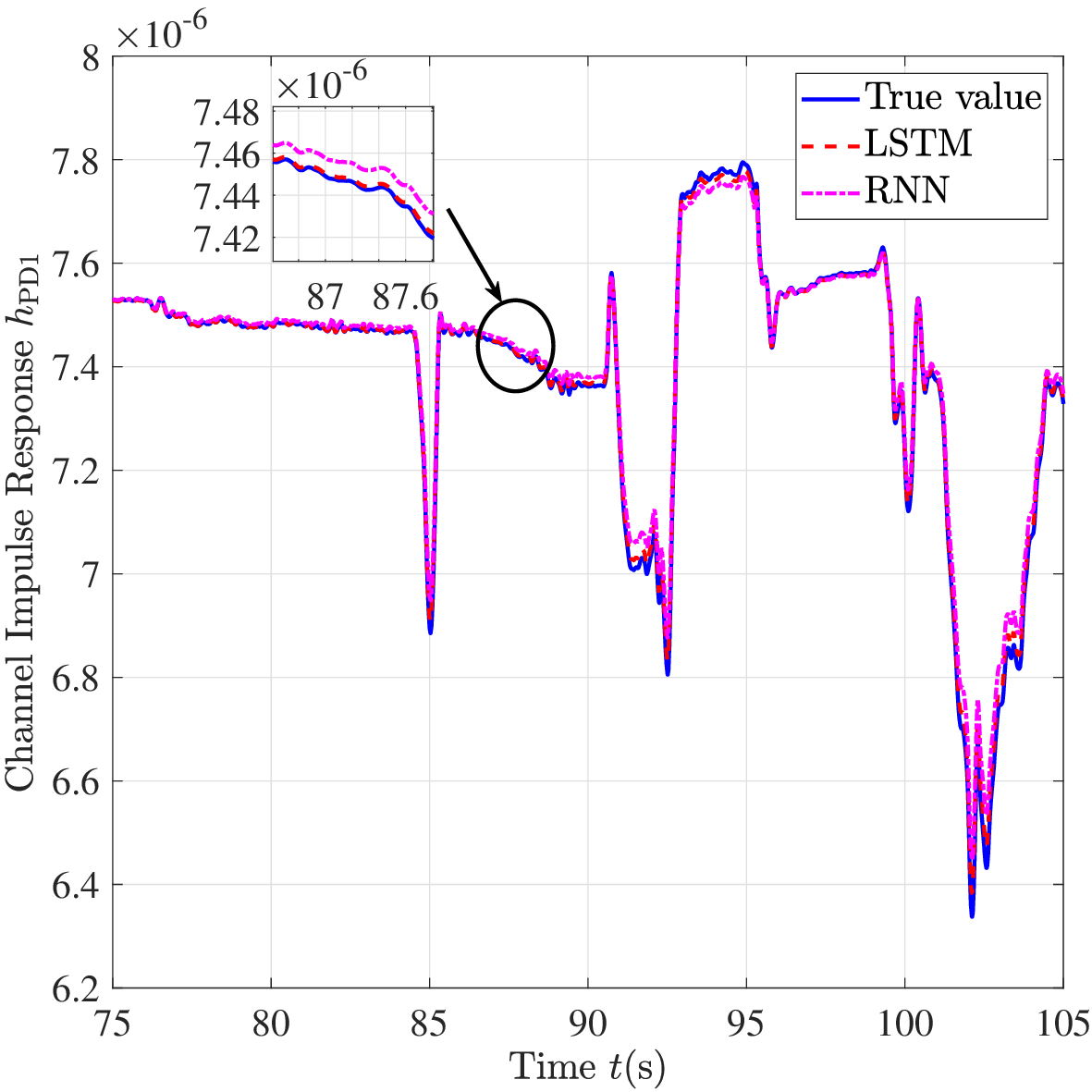}
		\vskip-0.2cm\centering {\footnotesize (a)}
	\end{minipage}\hfill
	\begin{minipage}[b]{0.45\textwidth}
		\centering
		\includegraphics[scale=0.35]{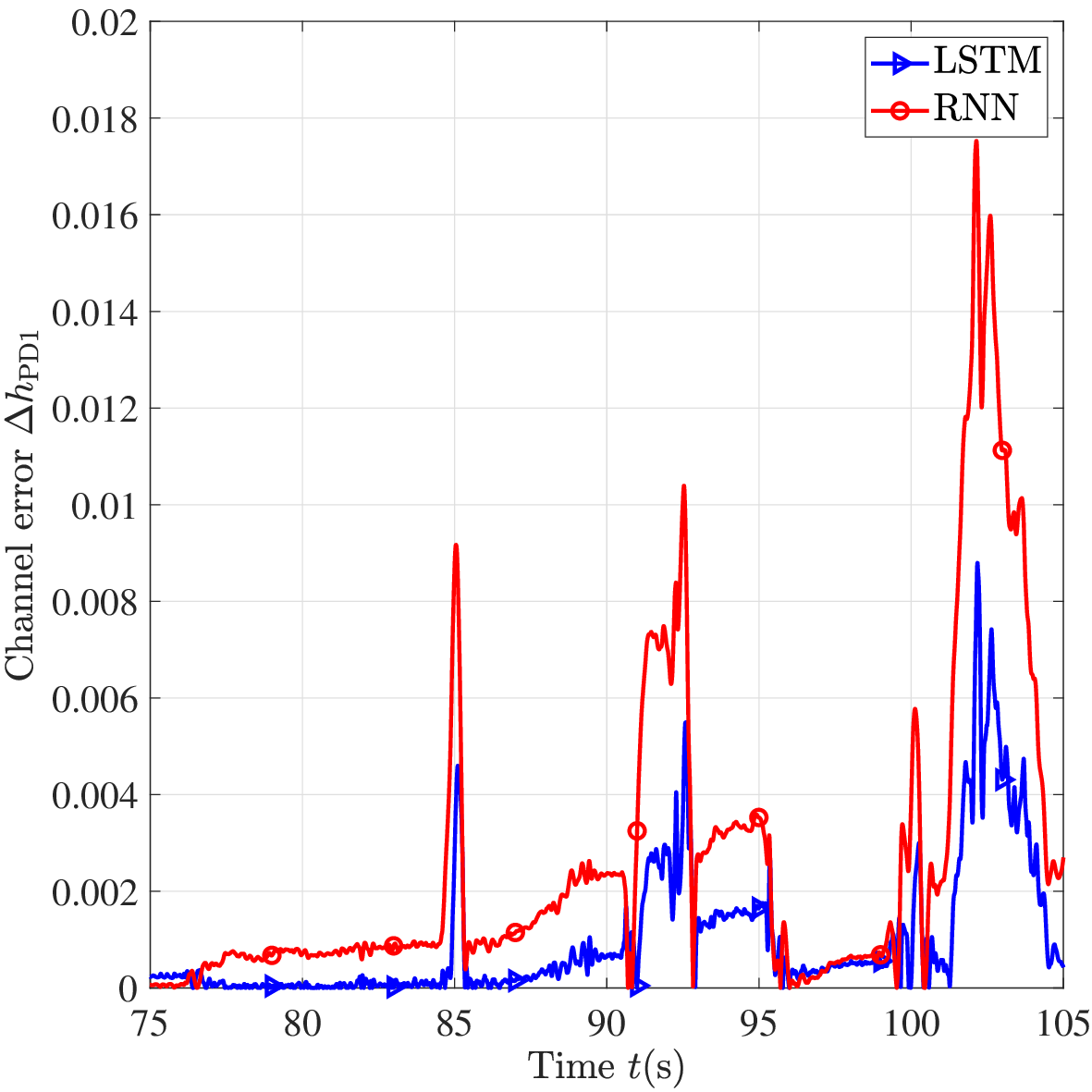}
		\vskip-0.2cm\centering{\footnotesize (b)}
	\end{minipage}\hfill
	\caption{Channel tracking diagrams: (a) Channel impulse response versus the time slot $t$, (b) channel error  versus the time slot $t$.}
	\label{hat}  
\end{figure}

We conducted the experiments to evaluate the channel tracking performance. To be more specific, we used PD1 to collect data in sitting scenarios with ${{\bf{u}}_l} = {\left[ {0,0,3} \right]^{\rm{T}}}$ and ${{\bf{u}}_u} = {\left[ {0,0,1} \right]^{\rm{T}}}$. A total of 53216 points were used, the first 37251 points were used for offline training and the last 15965 points were used for online testing. The parameters of the LSTM module are set as follows: the number of iterations is 150, the hidden size is 100, the time step is 4, and the learning rate is 0.01.

We compare the channel tracking performance of the proposed LSTM method and the RNN method  in Fig. \ref{hat}. Specifically, Fig. \ref{hat} (a) and Fig. \ref{hat} (b) show the channel tracking performance and the channel error results  versus the time slot $t$, respectively.
We can observe that for the time-varying channel, both the LSTM and RNN methods can realize the channel tracking function, and the former  has a better performance. The reason is that the LSTM is more suitable for long-term training compared to the RNN.


\section{conclusions}
Mobile LiFi is affected by time-varying channels. In this paper, we studied the real-time channel characteristics of LiFi system  by using the channel information obtained from practical measurements. Influencing factors of the mobile LiFi system are unveiled by the measured data.
Firstly, the concept of coherence time is adopted to obtain  accurate channel information of time-varying features. Our experimental results show that the coherence time  is on the order of tens of milliseconds, which is influenced by the number of PDs, the position of the person, and the walking speed.
Secondly, we derived the achievable data  rate expression to evaluate the communication performance as a function  of the PD number, PD position, and LED number. Experimental results show that the system can support a data rate
of $36 {\rm{Mbit/s}}$, while the application of multiple LEDs improves the performance even further.
Moreover, considering the performance of mobile LiFi, the channel  estimation coding and CDRN channel estimation schemes are compared with the LS scheme. Our results show that the CDRN estimator has the best performance and potential in   mobile LiFi. Finally, we adopt the LSTM method to track the real-time channel information. Our results show that LSTM is more suitable for mobile LiFi real-time channel tracking by comparing with the recurrent neural network  method.

\bibliographystyle{IEEE-unsorted}
\bibliographystyle{IEEEtran}
\bibliography{refs0611}

\end{document}